\documentclass{aa}
\usepackage{graphicx}
\usepackage{bm}
\usepackage{times}
\usepackage{amsmath}
\usepackage{amssymb}
\usepackage{natbib}
\usepackage{color}
\usepackage[colorlinks,allcolors=blue]{hyperref}
\usepackage{verbatim}
\bibpunct{(}{)}{;}{a}{}{,}
\graphicspath{{./fig/, ./png/}}

\renewcommand{\bm}{\boldsymbol}

\newcommand{\uuu}{{\bm u}}

\newcommand{\gggg}{\mbox{\boldmath $g$} {}}
\newcommand{\AAA}{{\bm A}}

\newcommand{\BBB}{{\bm B}}

\newcommand{\JJJ}{{\bm J}}

\newcommand{\UUU}{{\bm U}}

\newcommand{\OO}{\bm{\Omega}}

\newcommand{\Equ}[1]{Equation~(\ref{#1})}

\newcommand{\EQ}{\begin{equation}}
\newcommand{\EN}{\end{equation}}
\newcommand{\EQA}{\begin{eqnarray}}
\newcommand{\ENA}{\end{eqnarray}}

\newcommand{\dert}[1]{\frac{\partial #1}{\partial t}}
\newcommand{\Dert}[1]{\frac{{\rm D} #1}{{\rm D} t}}

\newcommand{\cP}{c_{\rm P}}
\newcommand{\cV}{c_{\rm V}}

\newcommand{\urms}{u_{\rm rms}}

\newcommand{\chiSGS}{\chi_{\rm SGS}}

\newcommand{\Co}{{\rm Co}}

\newcommand{\PraSGS}{{\rm Pr}_{\rm SGS}}

\newcommand{\Pm}{{\rm Pr}_{\rm M}}

\newcommand{\Rey}{{\rm Re}}

\newcommand{\Rm}{{\rm Rm}}

{}

\newcommand{\FFFrad}{{\bm F}^{\rm rad}}
\newcommand{\FFFSGS}{{\bm F}^{\rm SGS}}

\newcommand{\Figu}[1]{Figure~\ref{#1}}

\begin{document}

\authorrunning{Viviani \& K\"apyl\"a}

\titlerunning{Kramers opacity law and dynamo transitions}

  \title{Physically motivated heat conduction treatment 
in simulations of solar--like stars: effects on dynamo
transitions}

 \author{M. Viviani \inst{1,2}
          \and
          M. J. K\"apyl\"a \inst{3,2,4} 
        }
 
\institute{Dipartimento di Fisica, Universit{\`a} della Calabria, 
           I-87036, Rende (CS), Italy \\
           \email{viviani@mps.mpg.de} 
    \and
       Max Planck Institute for Solar System Research,
       Justus-von-Liebig-Weg 3, D-37077 G\"ottingen, Germany
    \and
      Department of Computer Science, 
      Aalto University, PO Box 15400, FI-00076 Aalto, Finland
    \and
    Nordita, KTH Royal Institute of Technology and Stockholm University, 
              Roslagstullsbacken 23, SE-10691 Stockholm, Sweden
}

\date{Received  / Accepted }

\abstract{
Results from global magnetoconvection simulations of solar-like stars are at odds with
observations in many respects: 
They show a surplus of energy in the kinetic power spectrum
at large scales, anti-solar differential rotation profiles, 
with accelerated poles and a slow equator, for the solar rotation rate, 
and a transition 
from axi-- to non--axisymmetric dynamos at a much lower 
rotation rate than what is observed.
Even though the simulations reproduce the 
observed active longitudes in fast rotators, 
their motion in the rotational frame (the so--called azimuthal dynamo wave, ADW) 
is retrograde, in contrast to the prevalent prograde motion 
in observations.
}
{
We study the effect of a more realistic treatment of heat conductivity in 
alleviating
the discrepancies
between observations and simulations.
}
{We use physically--motivated heat conduction, 
by applying Kramers opacity law, 
on a semi--global spherical setup
describing convective envelopes of solar-like stars,
instead of a prescribed heat conduction profile 
from mixing--length arguments.
}
{We find that some aspects of the results now better correspond 
to observations: The axi-- to non--axisymmetric transition 
point is shifted towards higher rotation rates. 
We also find a change in the propagation direction of ADWs so 
that also prograde waves are now found.
The transition from anti--solar to solar--like rotation
profile, however, is also shifted towards higher rotation rates, 
leaving the models into an even
more unrealistic regime.
}
{
Although a Kramers--based heat conduction
does not help in reproducing the solar rotation profile,
it does help in the faster rotation regime, 
where the dynamo solutions now match better with observations.
}


\maketitle

\section{Introduction}\label{sec:KrIntro}
The solar surface differential rotation has been known
for a long time \citep{1630Sch,Carr1863}:
the equator completes a turn in around 25 days, 
while the poles take roughly 30 days.
Helioseismic inferences have allowed also to uncover
the subsurface rotation \citep{Schouea98},
and revealed
that the lines of constant 
angular velocity are radial. This was 
somewhat unexpected, as
in a uniform, incompressible flow, the Taylor--Proudman
theorem \citep{Ch61} states that the horizontal components of 
the velocity field cannot vary in the direction of the rotation
axis, and the flow is forced to move in vertical columns, 
in which case angular velocity contours
constant on cylinders would be observed.
Hence, the Sun is able to break the Taylor--Proudman
balance with some means.
Another surprising observational result came
from time-distance helioseismology 
\citep{HDS12}, which revealed a lack of 
power in the kinetic energy spectrum at large scales,
where the peak for giant cells should be located.
Such a peak would be expected from mixing--length theory 
\citep[MLT;][]{Vi53, BV58}:
in its original formulation, MLT predicts convection
at all possible scales, which would also correspond
to cells of the diameter of the entire convective layer. 
Also, more recent measurements \citep[e.g.,][]{Rincon2017}
suggest that supergranulation may indeed be the largest scale excited
in the Sun.

Theoretical explanations as to how the Sun breaks the
Taylor--Proudman balance include a "thermal wind", generating 
a clockwise meridional circulation pattern.
This circulation  results from a 
latitudinal temperature gradient, which is such that the pole is warmer 
than the equator by a small amount, of the order of a few Kelvin \citep{R89}. 
This temperature difference is 
comparable to
the error of current instruments, 
although \cite{ROR08} reported on an enhancement of 
$\sim 2.5~{\rm K}$ at the Sun's poles.
One possibility to explain such a temperature gradient theoretically 
is to argue for 
the importance of
turbulent effects, such as latitudinally anisotropic 
heat flux, which has been shown to be able to lead to a temperature 
difference of $\sim 4~{\rm K}$ 
\citep{KR95}.
Also, the presence of a weakly subadiabatic layer at the 
base of the convection zone has been shown to generate a 
thermal wind and sustain the necessary temperature gradient
in a mean--field hydrodynamic model \citep{Re05}.

Modelling efforts
of stellar convection in spherical or semi--spherical
shells still struggle in producing solutions
in which the Taylor-Proudman balance
is self--consistently broken, and thus still tend to
show cylindrical isocontours for the differential
rotation \citep[e.g.,][]{GSKM13,GYMRW14,KKB14,ABMT15}.
This is commonly interpreted to imply
too strong a rotational influence.
Moreover, such models 
are unable to 
reproduce an accelerated equator when using the solar
rotation rate \citep{GYMRW14, KKB14, KKKBOP15}.
Most of present--day numerical setups are using fixed heat conduction
profiles and depths of convection zones \citep[e.g.,][]{BMT11,KMCWB13}, 
motivated by MLT.
Although MLT--designed setups have been successful in reproducing the 
pattern of granulation and supergranulation in surface 
convection
\citep[e.g.,][]{BCNS05, NSA09},
models simulating deeper parts of the convection zone (CZ)
produce far more power 
in the velocity spectrum
at
large scale than observations \citep[][]{GB12}. 
All the above mentioned discrepancies between observations and
numerical models are collectively known as
``convective conundrum'' \citep[see, e.g.,][]{2016AdSpR..58.1475O}
and solving it is one of the major challenges of 
contemporary solar physics.

One proposed way to crack the convective
conundrum is to
hypothesize that the actual convectively unstable 
layer in the Sun, according to the Schwarzschild 
criterion \citep{Ch61}, is shallower than expected.
\cite{Sp97} described convection 
as being driven by cool threads descending 
from the surface into deeper layers,
overwhelming the convection driven by heating
from below.
Such a phenomenon is 
now denoted as \textit{entropy rain} \citep{Br16}
and describes surface--driven convection, that would excite 
only small to 
medium length scales.
By extending MLT to include entropy fluctuations, \cite{Br16} 
identified the presence of a Schwarzschild stable,
sub--adiabatic,
layer in which the convective flux is still positive.
Such a layer was identified first in the Earth's atmosphere
\citep{1961JAtS...18..540D, De66},
and hence was termed as ``Deardorff layer''.

The formation of such sub--adiabatic layers has been reported 
in the hydrodynamic studies
of, e.g., \cite{2017ApJ...843...52H},
\cite{2017arXiv170400817K}, and \cite{2017arXiv170306845K}.
Especially the 
study 
by \cite{2017arXiv170306845K}
is relevant here, as they demonstrated
the emergence of a substantial sub--adiabatic layer, and the 
existence of non--local surface driving of convection, by using
Kramers opacity law in a Cartesian model. 
Consequently, they redefined the convection zone as the sum 
of the convection zone in the traditional sense, now called
buoyancy zone, plus the sub--adiabatic part, now denoted as 
Deardorff layer.
The depth of the layers was not determined a priori, 
but rather was an outcome of the simulations.

However, these studies 
did not investigate large--scale dynamo action.
The effect of sub--adiabatic layers in global MHD simulations
was investigated by \cite{KVKBF19}.
The formation of a stably stratified layer
at the bottom of the domain allowed for the storage of 
magnetic field beneath it, also found in an earlier study 
by \citep[][]{BMBT06}, but these strong fields were also observed 
to be capable of suppressing the oscillating 
magnetic field at the surface. 
\cite{KVKBF19} also considered the effect of sub--adiabatic layers on 
the convective velocity spectra, but found that the decrease in 
power at large scales was not enough to solve 
this part of the conundrum.

Another mechanism to reduce the 
too high rotational influence on convection
in simulations, studied first in a Cartesian 
model by \cite{HRY15b} and then in fully 
spherical models by \cite{KMB18}, 
could be provided by
the Lorentz force feedback from the magnetic to the velocity field.
Such feedback could result from strong magnetic fluctuations, 
originating e.g., from the action of a 
small--scale dynamo instability \citep[see, e.g.,][]{1968JETP...26.1031K} 
operating the the CZ.
Thereby generated magnetic fluctuations 
could suppress the turbulent velocity field
through the Lorentz force, hence
acting as an enhanced viscosity, 
and increasing the magnetic Prandtl number, 
the ratio of the viscosity and resistivity of the fluid.
\cite{KMB18} investigated such a situation numerically, and
their simulations developed an overshoot zone at the base of the 
domain, and also showed a decrease in the convective power at large
scales, due to downward directed plumes.
These results, although arising for a different reason,
are consistent with the results of \cite{2017arXiv170306845K}
and \cite{KVKBF19}.
Another finding of \cite{KMB18} was that the plumes, carrying 
their angular momentum inward, caused the rotation profile to 
switch to anti--solar.

Observations of rapidly rotating stars, younger and more 
active than the Sun, indicate concentrations of 
magnetic activity at high latitudes persisting for a long time
\citep[e.g.,][]{BT98}.
A common configuration is two activity patches on two 
``active" longitudes, separated roughly by 180 degrees in longitude
\citep[e.g.,][]{Jetsu1996}.
Active longitudes usually migrate in the rotational frame of the star,
forming azimuthal dynamo waves (ADW) \citep[e.g.,][]{BT98}.
The direction of migration of these structures can follow the 
plasma rotation, and in this case we will talk about prograde
ADWs; they can also drift in the opposite direction (retrograde ADWs),
or they can stand still with respect to the observer's point of view
(standing ADW).
These ADWs can persist for time spans extending to ten years
\citep[e.g.,][]{Marjaana11}, or their appearance can be more 
erratic \citep[e.g.,][]{Olspert15}, with a short--lived ADW 
reappearing after some time.
\cite{Lehtinen16} and \cite{2016MNRAS.462.4442S} reported on a
threshold in activity, above which stars show active longitudes. 
In the study of \cite{Lehtinen16}, the active longitudes 
found were mostly migrating in the prograde direction.
The appearance of active longitudes has been attributed to
non--axisymmetric dynamo modes operating in these stars 
\citep{TBK02}, in contrast to the axisymmetric dynamo operating 
in less active stars.
The transition from non--axisymmetric to axisymmetric dynamos
has also been studied numerically \citep{CKMB14, VWKKOCLB18}, 
but these studies reported a majority of retrograde 
ADWs, and a transition from axi-- to non--axisymmetric solutions 
at too low rotation rates in comparison to observations.
Both studies were using prescribed and MLT-motivated profiles 
for heat conduction, resulting in a priori fixed depth of the 
convection zone.

The aim of this paper is to extend  the study of 
\cite{VWKKOCLB18} to include a dynamically adaptable heat 
conduction. 
In order to do this, we use a Kramers--like opacity law, as was done in
\cite{KVKBF19} for semi--spherical wedge simulations.
We use computational domains extending over the full 
longitudinal extent to be able to study both axi-- and 
non--axisymmetric dynamo solutions.

\section{Setup and Model}\label{sec:KrModel}
We apply a similar setup as in \cite{KMCWB13} and \cite{KVKBF19}, 
representing the outer envelope of a solar-like star,
$0.7R\leq r\leq R$ (with $R$ the radius of the star),
in a semi--spherical domain, $0\leq \phi \leq 2\pi$
and $\theta_0 \leq \theta\leq \pi - \theta_0 $
($\theta_0 = 15^o$).
We solve numerically the system of MHD equations:
\begin{equation}\label{eq:Krmodel}
\begin{aligned}
\Dert{ \rm ln \rho} &+ \nabla \cdot \UUU = 0, \\
\Dert{\UUU} &= \gggg - 2 \OO_0 \times \UUU + \frac{1}{\rho} 
                      \left( \JJJ \times \BBB - \nabla p + 
                      \nabla \cdot 2\nu\rho \bm{\mathsf{S}} \right), \\
T\Dert{s} &= \frac{1}{\rho} \left[ - 
                  \nabla \cdot \left( \FFFrad + \FFFSGS \right) + 
                   \mu_0 \eta \JJJ^2 -\Gamma_{\rm cool}\right] + 2\nu \bm{\mathsf{S}}^2, \\    
\dert{\AAA} &= \UUU \times \BBB - \mu_0 \eta \JJJ, \\
\end{aligned}
\end{equation}
where $\rho$ and $\UUU$ are the density and the velocity field,
$\gggg=-GM/r^3$ is the gravitational acceleration, with $G$ being the
gravitational constant and $M$ the mass of the star.
$\OO_0=\Omega_0 \left( \cos \theta , - \sin \theta, 0 \right)$ 
is the bulk rotation.
$\JJJ$, $\BBB$ and $\AAA$ are the electric current, the magnetic field 
and the vector potential, respectively, $p$, $\nu$ and $\mu_0$
are the pressure, the viscosity, and the magnetic permeability in
vacuum, while $\eta$ is the magnetic diffusivity.
$\bm{\mathsf{S}}$ is the rate--of--strain tensor.
$\FFFrad$ and $\FFFSGS$ are the radiative and sub--grid scale (SGS) 
fluxes, expressed by:
\begin{equation}\label{eq:KrFluxes}
    \FFFrad = -K \nabla T \quad
    \FFFSGS = -\chiSGS\rho T\nabla s' \quad.
\end{equation}
$K$ is the radiative heat conductivity and $\chi_{\rm SGS}$ is
the SGS heat diffusivity, assumed to be constant.
$s'$ is the fluctuating entropy,
$s'=s-\left\langle \overline{s}\right\rangle_{\theta}$,
where the overbar denotes longitudinal average and the 
brackets express averaging over the variable in the subscript.
Finally, $\Gamma_{\rm cool}$ is a term acting near the surface
and cooling towards a reference temperature.
Its flux is expressed as:
\begin{equation}\label{eq:KrGammacool}
    \overline{F}_{\rm cool}=\int_{0.7R}^{R} \Gamma_{\rm cool}dr.
\end{equation}

The initial velocity and magnetic fields are gaussian seeds.
The initial stratification is isentropic.
The radiative heat conductivity, $K$, follows from Kramers opacity law 
for free--free and bound--free transitions
\citep[used also in][]{BB14,2017arXiv170306845K,K18,KVKBF19,KGOKB19}:
\begin{equation}\label{eq:KrKheat}
    K = K_0 \left( \frac{\rho}{\rho_0}\right)^{2}
                  \left( \frac{T}{T_0}\right)^{13/2},
\end{equation}
where $\rho_0$ and $T_0$ are reference values for density 
and temperature.
The constant $K_0$ is defined via:
\begin{equation}
\begin{aligned}
    K_0 &= \frac{\mathcal{L}}{4\pi} c_v \left(\gamma -1 \right)
              \left( n_{\rm ad} +1\right)\rho_0\sqrt{GMR}, \\
        \mathcal{L} &= \frac{L_0}{\rho_0 \left( GM\right)^{3/2}R^{1/2}}.
\end{aligned}        
\end{equation}
$\mathcal{L}$ is the normalized luminosity, $\cV$ is the specific 
heat at constant volume, $\gamma=\cP/\cV$ is the ratio between the
specific heat at constant pressure and 
volume, and $n_{\rm ad}=1.5$ 
is the adiabatic index.

The velocity field is impenetrable and stress free at 
all boundaries, while entropy derivatives are set 
to zero at $\theta = \theta_0$ and $\theta=\pi-\theta_0$.
The magnetic field is radial at $r=R$ and a perfect 
conductor boundary condition is applied at the bottom boundary.
At the latitudinal boundaries, $\BBB$ is tangential,
which means, in terms of the vector potential:
\begin{equation}
    A_r = A_{\phi} = \frac{\partial A_{\theta}}{\partial \theta}=0
    \qquad {\rm at} \quad \theta = \theta_0 \quad {\rm and}
    \quad \theta=\pi-\theta_0
\end{equation}
\cite{KGOKB19} showed that this latitudinal boundary condition
does not generate major differences with respect to the 
perfect conductor boundary condition used in previous works
\citep[e.g., in][]{KMCWB13,CKMB14,WKKB14,VWKKOCLB18}.

The simulations are defined by the parameters 
$\Omega_0$, $\nu$, 
$\eta$,
$\chi_{\rm SGS}$, $K_0$, 
$\rho_0$, $T_0$ and the energy flux at the bottom, 
$F_{\rm bot}=-K\partial_r T|_{r=0.7R}$. 

Moreover, important non--dimensional input parameters 
are the magnetic and SGS Prandtl numbers:
\begin{equation}\label{eq:KrPrandtl}
    \Pm = \frac{\nu}{\eta}, \quad    
    \PraSGS = \frac{\nu}{\chiSGS}.
\end{equation}
Output parameters of the simulations are the fluid and magnetic Reynolds numbers:
\begin{equation}\label{eq:KrRey}
    \Rey = \frac{\urms}{\nu k_f}, \quad 
    \Rm = \frac{\urms}{\eta k_f}
\end{equation}
with $\urms=\sqrt{3/2 \left\langle U_r^2 + U_{\theta}^2 
                                      \right\rangle_{r\theta\phi t} }$
rms velocity and $k_f=2\pi/0.3R$ the wavenumber
of the largest eddy, corresponding to the radial extent,
and the Coriolis number:
\begin{equation}\label{eq:KrCo}
    \Co = \frac{2\Omega_0}{\urms k_f},
\end{equation}
quantifying the relative importance of rotation and convection.

Physical units are chosen using the solar radius 
$R=7\cdot 10^8~{\rm m}$, the solar angular velocity
$\Omega_\odot=2.7\cdot 10^{-6}~{\rm s^{-1}}$,
the density at the bottom of the solar convection zone
$\rho_{\rm bot}=200~{\rm Kg~m^{-3}}$, and the magnetic permeability
in vacuum $\mu_0 = 4\pi \cdot 10^{-7}~{\rm H~m^{-1}}$.
We performed our simulations using the 
{\sc Pencil Code}\footnote{\url{https://github.com/pencil-code/}},
a high--order, finite--difference, open source code 
for solving the magnetohydrodynamic equations.

\section{Results}\label{sec:KrRes}
The simulations and their defining parameters are 
summarized in Table~\ref{tab:Krruns}.
Run~R1 and R2 correspond to
Run~C3
and D
of \cite{VWKKOCLB18}, where 
the radiative heat conductivity K was only a function of depth,
as described in \cite{KMCWB13}.
Here, our aim is to study the effect of 
the more physical treatment for heat
conduction on the anti--solar to solar--like differential rotation
transition and the transition to non--axisymmetric magnetic fields.
Run~C3 was the simulation with the slowest rotation rate
showing both accelerated equator
and non-axisymmetric magnetic field,
hence it is a good choice for this study.
Run~D had a rotation rate 2.1 times the solar value, exhibited a 
non-oscillatory dynamo solution with dominance of the $m=1$ Fourier 
mode with a retrograde ADW; Run~C3 behaved otherwise similarly, but 
the dynamo solution was oscillatory. 
This difference in between the runs 
was most likely connected to 
the higher amount of differential rotation in C3 than in D, 
see also Sect.\ref{sec:appendix}, where we reproduce the 
rotation profiles of these runs. 
Run~R3 is the extension of the three times solar rotation rate 
Run~MHD2 of \cite{KVKBF19} over the full longitudinal extent.
Run~MHD2 was a wedge simulation covering 1/4 of the full longitude,
hence not allowing for non--axisymmetric solutions to develop.
We repeat this run in an extended azimuthal domain to study the 
possible topological changes of the magnetic field. 
Run~R4 has the same setup as Run~R3, but twice the rotation rate.
Simulations in the same rotation range, but with fixed heat conduction
profiles \citep[e.g.,][]{VWKKOCLB18}, all showed a clear dominance
of the non--axisymmetric mode over the axisymmetric one, and ADWs with 
retrograde migration.

\begin{table*}[h!]
\centering
\begin{tabular}{c | c c c | c c c c c c c |r r r}
 \hline
 Run & $\Omega [\Omega_\odot ]$ & $\PraSGS$ & $\Pm$ & $\Rey$ & $\Rm$ & $\Co$ & $\Co_{\rm rev}$ & rBZ & rDZ & rOZ & $\Delta_r$ & $\Delta_{\theta}$ \\
 \hline\hline
 R1 & 1.8 & 0.33 & 1.0 & 32 & 32 & 2.7 & 2.1 & 0.769 & 0.767 & 0.710 & -0.05 & -0.07\\ 
 R2 & 2.1 & 3.00 & 1.0 & 20 & 20 & 3.6 & 2.29 & 0.834 & 0.809 & 0.784 & -0.09 & -0.17\\ 
 R3 & 3.0 & 1.0 & 1.0 & 29 & 29 & 4.2 & 3.7 & 0.773 & 0.738 & 0.706 & 0.02 & 0.03 \\ 
 R4 & 6.0 & 1.0 & 1.0 & 24 & 24 & 10.2 & 8.8 & 0.778 & 0.740 & 0.710 & 0.03 & 0.04\\ 
 \hline
\end{tabular}
\caption{Summary of the runs. 
Columns 9--11
indicate the latitudinally averaged values for 
the depths of BZ, DZ and OZ.
Last two columns express the values for radial and latitudinal
differential rotation, as defined in 
Section~\ref{subsub:DR}.
}
\label{tab:Krruns}
\end{table*}

\subsection{Convection zone structure}\label{sub:Conv}
\begin{figure*}[ht!]
\begin{center}
\includegraphics[width=0.23\textwidth]{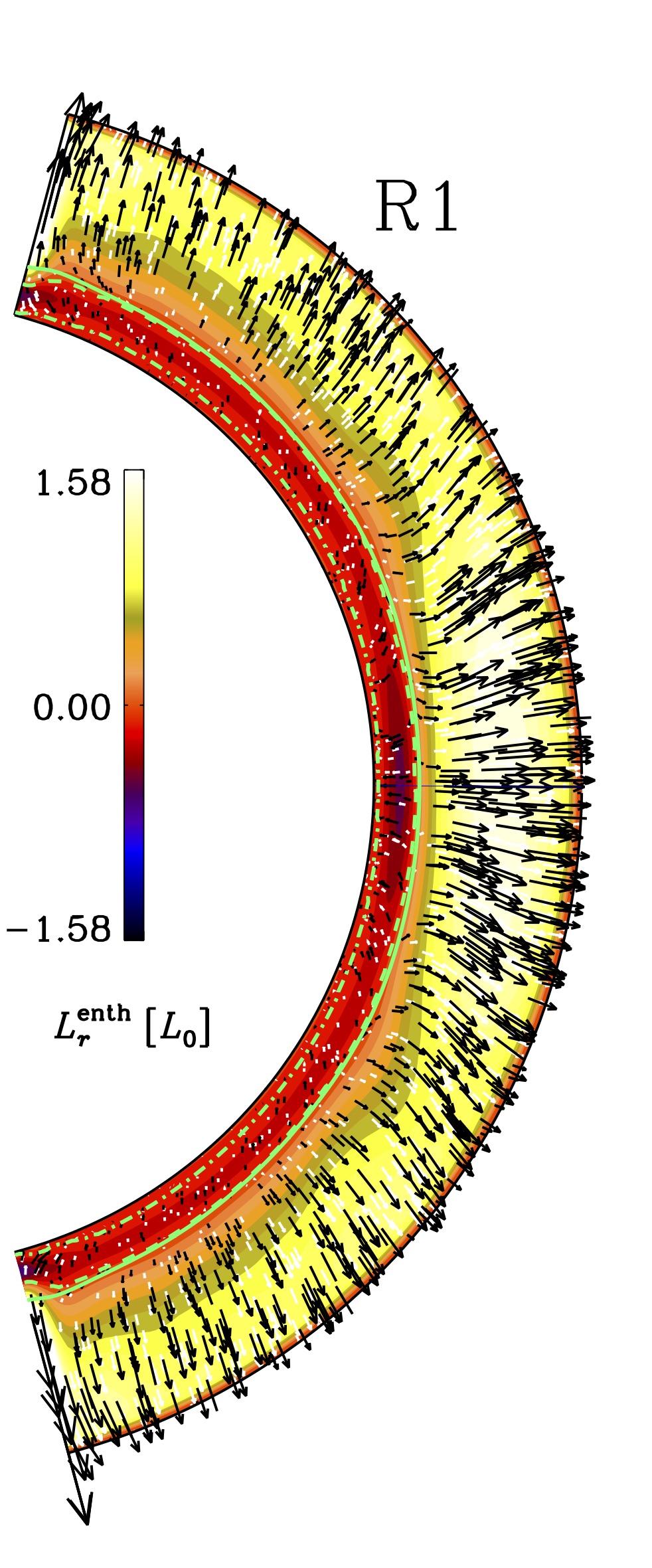}
\includegraphics[width=0.23\textwidth]{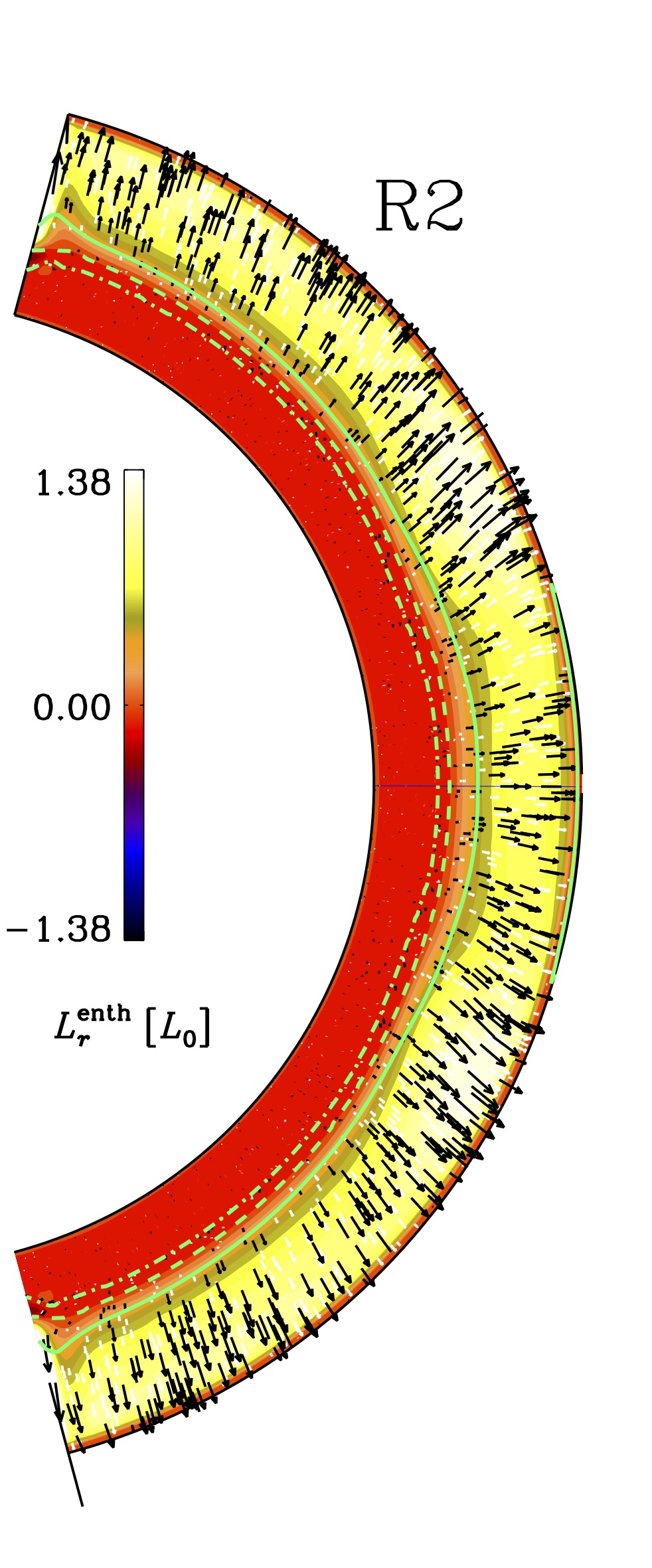}
\includegraphics[width=0.23\textwidth]{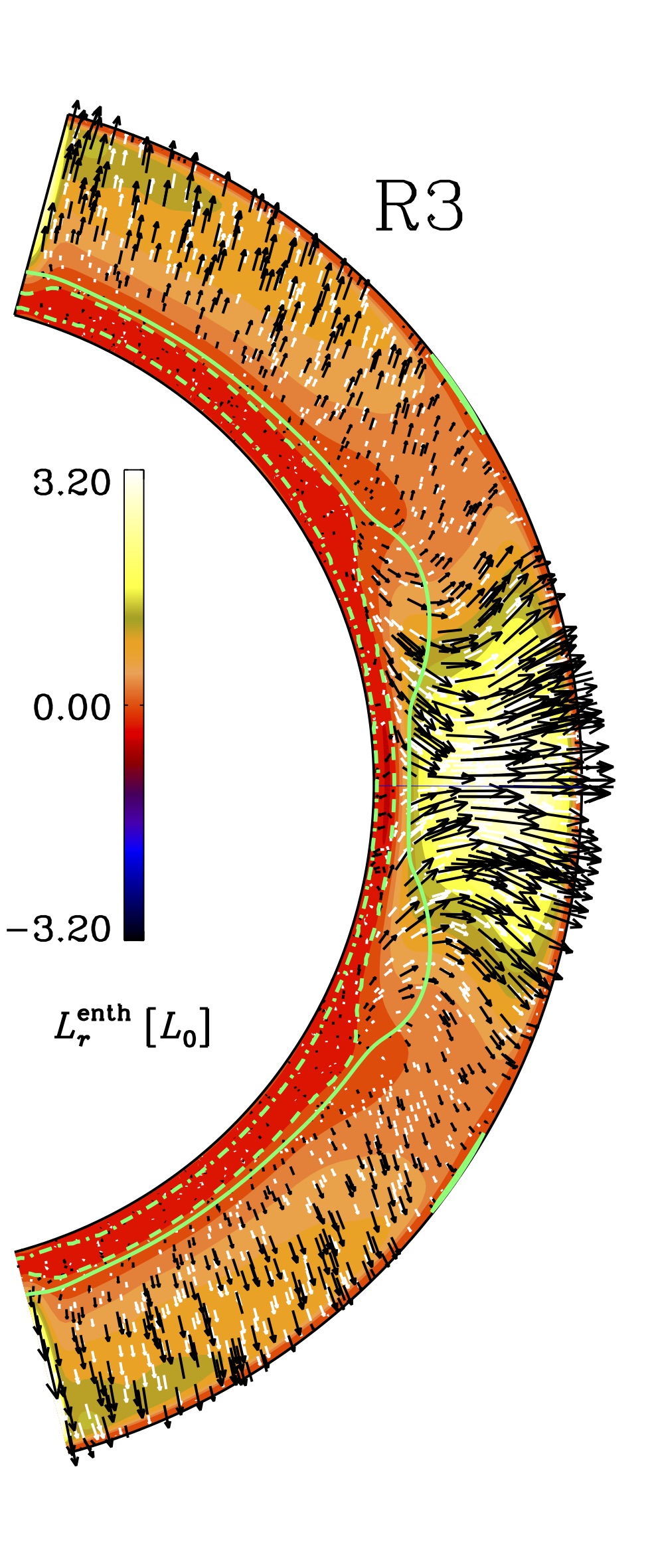} 
\includegraphics[width=0.23\textwidth]{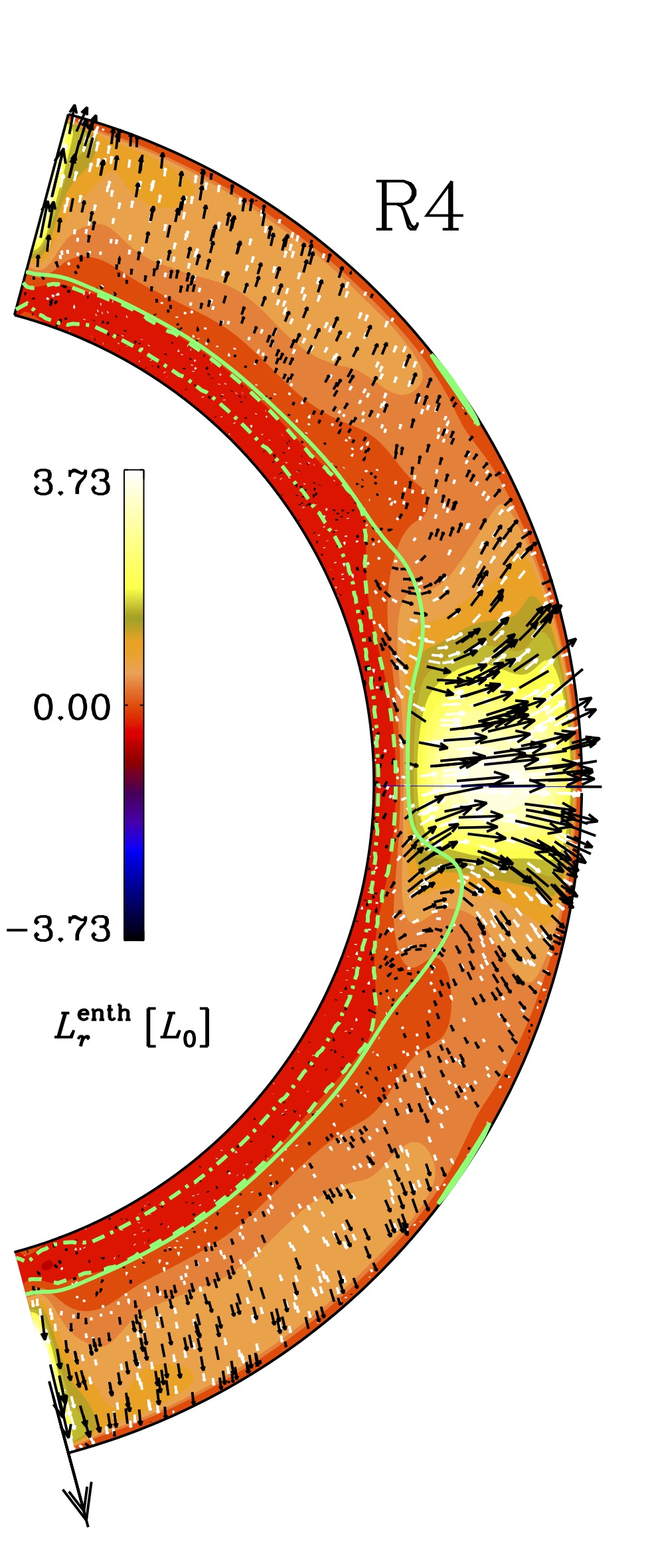}

\end{center}\caption{
Radial $L^{\rm enth}$ normalized by $L_0$.
The arrows show the direction of $L^{\rm enth}$.
The continuous, dashed and dash-dotted green lines represent,
respectively, the bottom boundaries of BZ, DZ and OZ.
}\label{fig:KrpFenth}
\end{figure*}

We define the convection zone according to the revised structure 
proposed by \cite{KVKBF19}, and indicate the bottom of the different
layers in Figures~\ref{fig:KrpFenth} and \ref{fig:KrpOm}
and Figure~\ref{fig:KrComp} in Appendix~\ref{sec:appendix}.
The radial enthalpy flux is defined as
$F^{\rm enth}_r =\overline{c_p \left(\rho \uuu_r\right)' T}$.
The bottom of the buoyancy zone (BZ), in which the radial enthalpy
flux is greater than zero, $F_r^{\rm enth}>0$, and the radial 
entropy gradient is negative, $\partial_r s <0$, is indicated 
with a continuous green
line; we note that our BZ would be the 
convection zone if defined based on the Schwarzchild criterion.
We denote the bottom of the Deardorff layer (DZ),
in which $F_r^{\rm enth}>0$ and $\partial_r s >0$, 
by a dashed line, and the bottom of the overshoot zone (OZ),
for which $F_r^{\rm enth}<0$ and $\partial_r s >0$, with a
dash--dotted line.
What we denote as the convection zone is the combination of the BZ and DZ,
where enthalpy flux is positive, but entropy gradient can be also positive,
meaning that the DZ part of our convection zone is sub--adiabatic.
In the radiative zone (RZ), $F_r^{\rm enth}\approx 0$ and 
$\partial_r s >0$.
The values averaged over latitude and longitude
for the depth of the layers are also shown in Table~\ref{tab:Krruns}.
We quote two Coriolis numbers for each simulation.
The first one is obtained from \Equ{eq:KrCo};
the second one,
denoted as $\Co_{\rm rev}$
in Table~\ref{tab:Krruns}, 
takes into consideration the wavenumber of the revised convection 
zone (BZ and DZ),
therefore we use $k_{f{\rm rev}}=2\pi/\left(R-r_{\rm DZ} \right)$
\citep[see, also,][]{KVKBF19},
where $r_{\rm DZ}$ is the latitudinally averaged radius of the
Deardorff layer,
reported in Table~\ref{tab:Krruns}.

Run~R1 has the deepest BZ of all runs,
a very thin DZ and a considerable OZ.
A thin RZ develops at the bottom.
Run~R2 has the thinnest convection zone of all runs. 
Also the OZ is thin, hence the run develops a very thick RZ.
In Runs~R1 and R2 the thickness of the layers does
not change considerably as function of latitude,
except
for a slight tendency of the DZ 
of becoming
thicker near the equatorial 
region for Run~R2. 
For Run~R3 the convection zone structure at higher latitudes 
again resembles the one of Run~R1. In the equatorial region, 
however, the convection zone becomes very deep. 
Close to the tangent cylinder the BZ becomes considerably shallower, 
and the DZ developes a "bulge" in that region. 
Run~R4 also exhibits a convection zone structure that
varies strongly with latitude, and closely resembles the one seen in Run~R3.
Also a hemispheric asymmetry develops in Run~R4: 
the DZ "bulge" is larger and the BZ is deeper in the lower hemisphere than 
in the upper one.

\subsubsection{Enthalpy flux}\label{subsub:fluxes}
We inspect the radial enthalpy flux, $F_r^{\rm enth}$,
by representing the enthalpy luminosity, 
$L^{\rm enth}=4\pi r^2F_r^{\rm enth}$, in
Figure~\ref{fig:KrpFenth} with black arrows.
The enthalpy flux in Run~R1 
and R2
is isotropic in latitude and rather 
radial everywhere in the BZ.
There is a slight tendency of the flux being 
enhanced/decreased in the equatorial region for Run~R1/R2.
A weak negative flux in the
equatorial region is present in the OZ
for Run~R1.
A different situation arises for the two 
more rapidly rotating runs,
where the convective transport of energy is stronger at low
latitudes.
Especially for Run~R3, there is a decrease of the enthalpy 
flux in the regions of the tangent cylinder.
A clear equatorial asymmetry is present in $L_r^{\rm enth}$ 
for Run~R4.
This asymmetry is also reflected 
in the convection zone structure, as discussed above.

\subsubsection{Differential rotation}\label{subsub:DR}

\begin{figure*}[ht!]
\begin{center}
\includegraphics[width=0.23\textwidth]{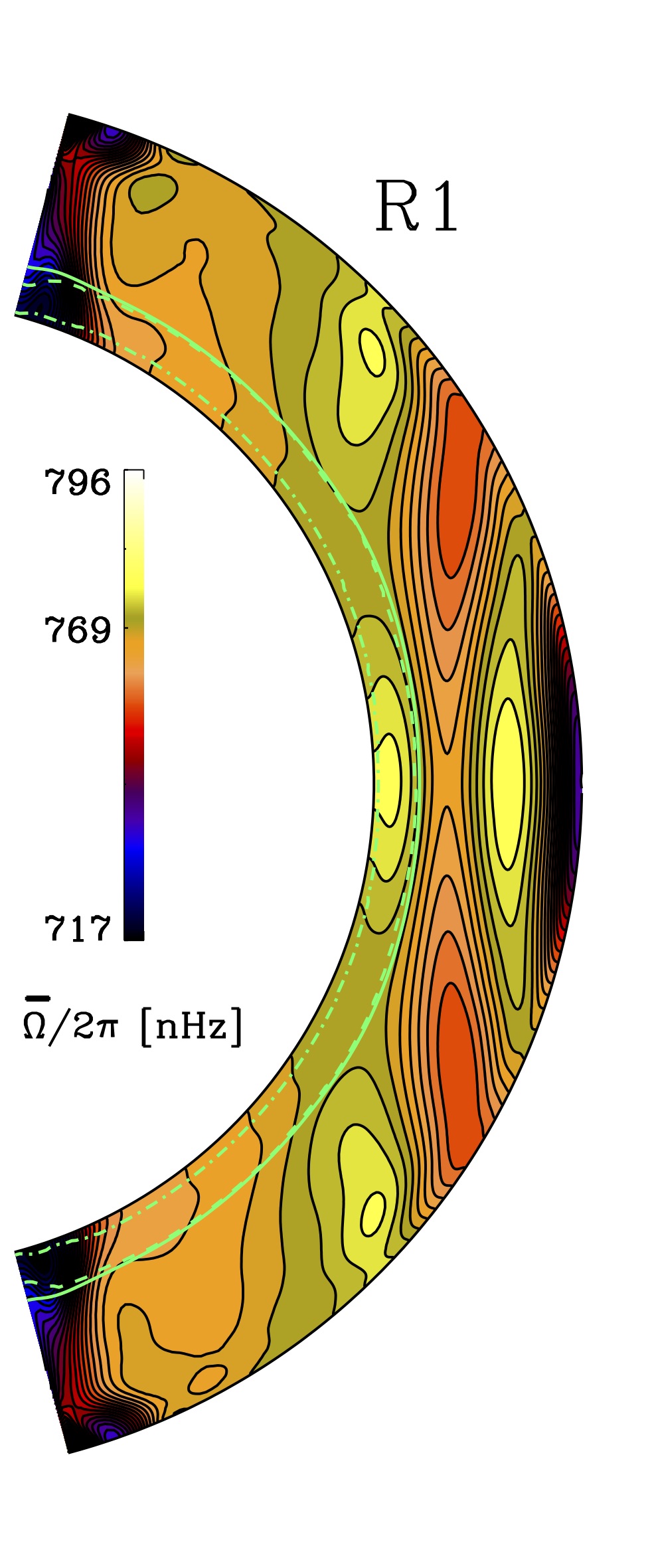}
\includegraphics[width=0.23\textwidth]{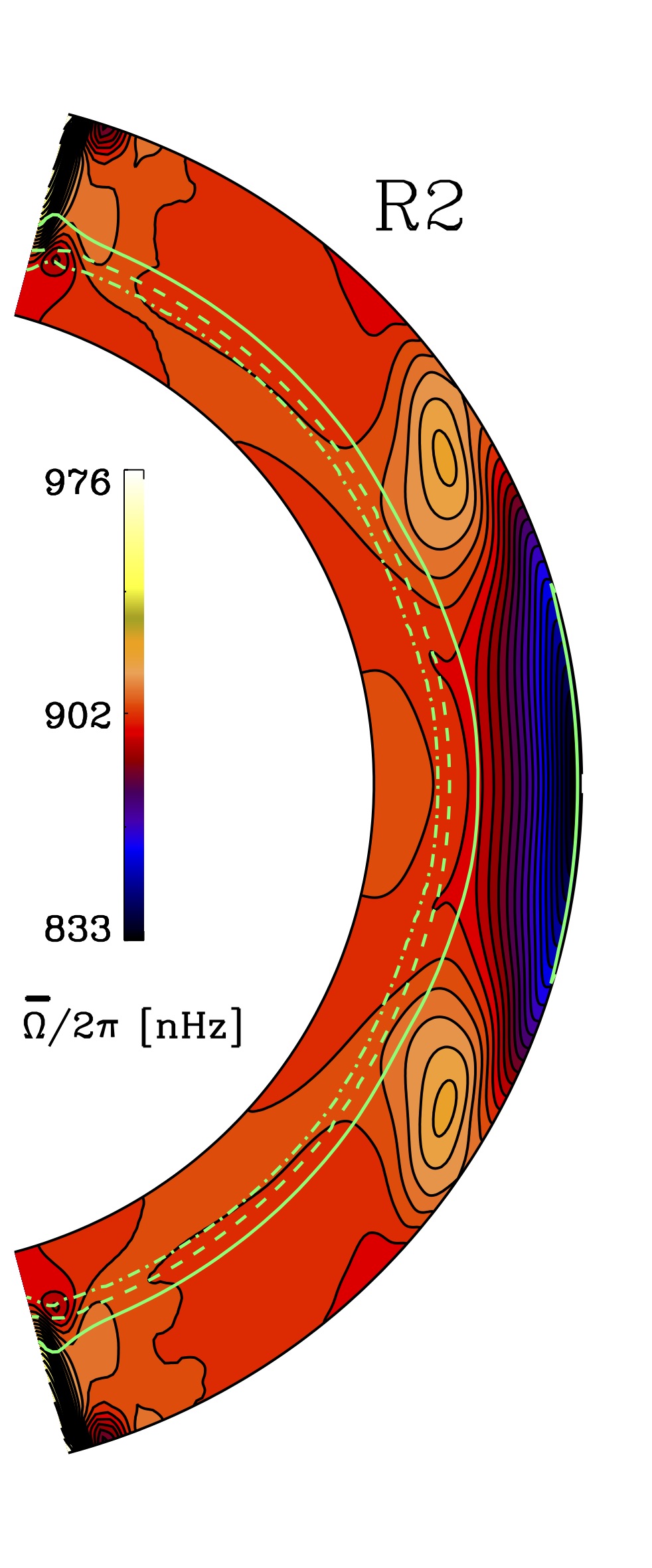} 
\includegraphics[width=0.23\textwidth]{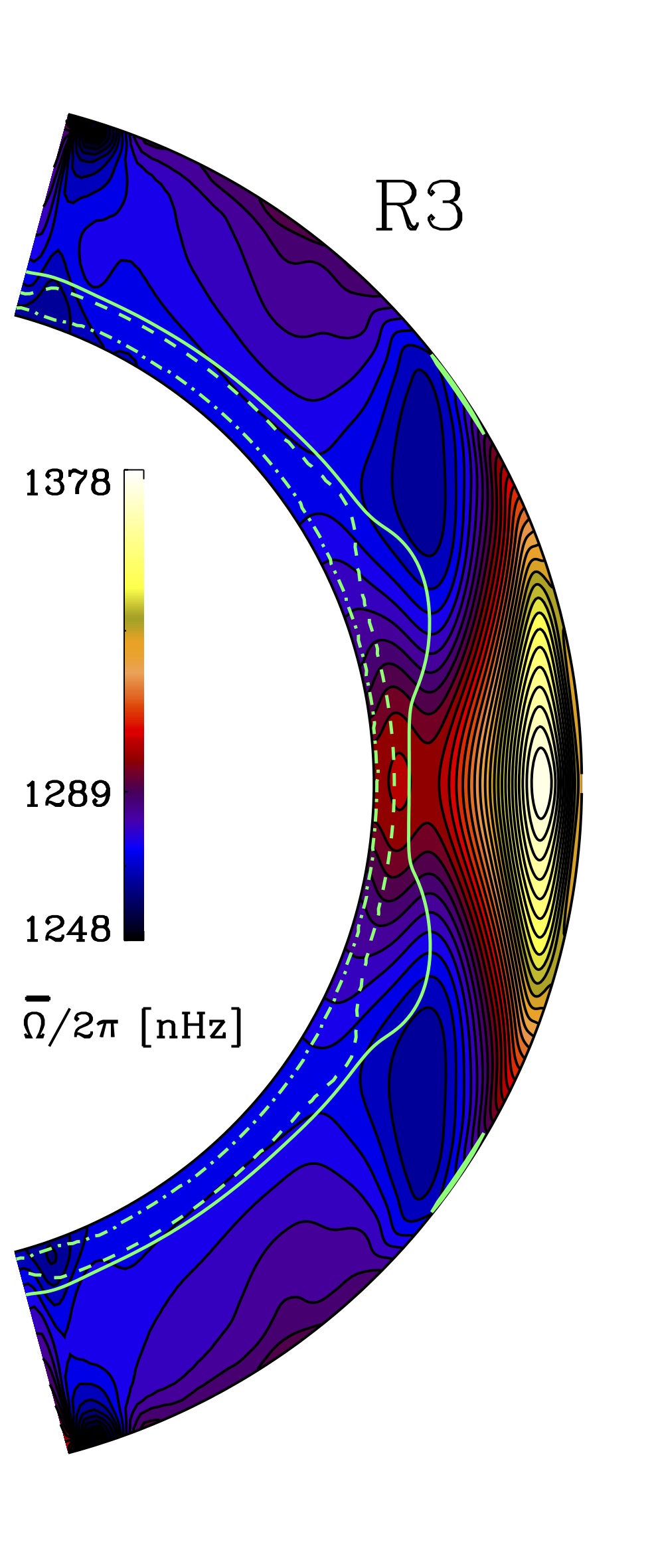}
\includegraphics[width=0.23\textwidth]{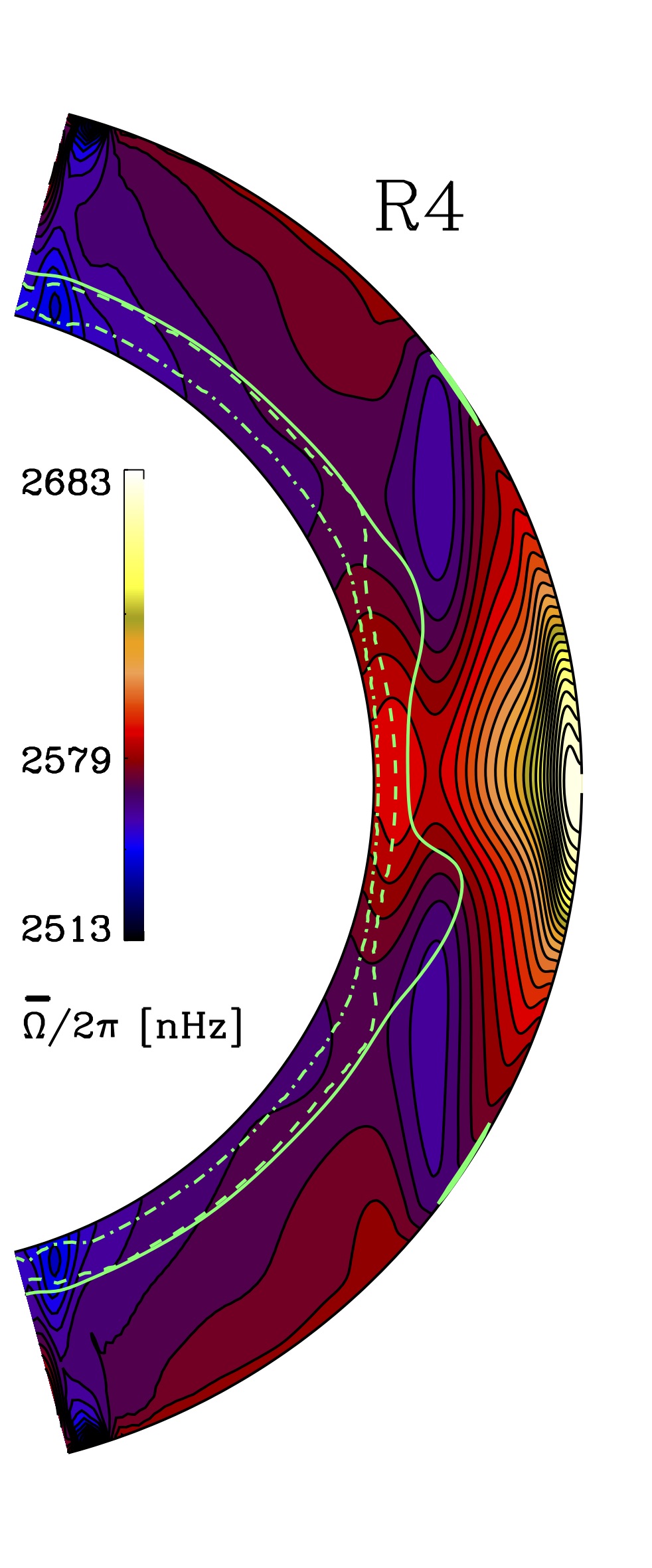}

\end{center}\caption{Differential rotation profiles.
Continuous, dashed and dash--dotted green lines as in
Figure~\ref{fig:KrpFenth}.}\label{fig:KrpOm}
\end{figure*}

The last two columns in Table~\ref{tab:Krruns} quantify the relative
radial and latitudinal differential rotation, defined as:
\begin{equation}
    \Delta_r = \frac{\Omega_{\rm eq} - \Omega_{\rm bot}}{\Omega_{\rm eq}}
    \quad {\rm and} \quad
    \Delta_{\theta} = \frac{\Omega_{\rm eq} - \Omega_{\rm pole}}{\Omega_{\rm eq}}.
\end{equation}
Here, $\Omega_{\rm eq}$ is the surface rotation rate at the equator,
$\Omega_{\rm bot}$ the equatorial rotation rate at the bottom of the
simulation domain
and $\Omega_{\rm pole}=\left( \Omega(R,\theta_0) + \Omega(R,\pi-\theta_0) \right)/2$.
We show the differential rotation profiles in Figure~\ref{fig:KrpOm} 
and compare with the corresponding simulations from other works, 
showing their profiles in Appendix~\ref{sec:appendix}.

Run~R1 corresponds to the simulation with the lowest rotation 
rate showing an accelerated equator in \cite{VWKKOCLB18}, the 
rotation profile in that case being quite solar--like (see
Figure~\ref{fig:KrComp}, first panel).
With an adaptable heat conduction prescription
the rotation profile is less
solar--like (see, \Figu{fig:KrpOm}, left panel
and Table~\ref{tab:Krruns}):
the equatorial acceleration becomes less pronounced,
the angular velocity contours are more cylindrical, and
additional regions of negative shear appear at mid--latitudes 
and in the equatorial region close to the surface. 
Such regions of negative shear at mid--latitudes, 
in many simulations \citep[e.g.,][]{KMB12,WKKB14},
have been identified as
responsible for the equatorward propagation of the magnetic
field at the surface.
In this case, however, we measure weaker relative
differential rotation and also the opposite sign in 
comparison to the results of \cite{VWKKOCLB18}, 
where $\Delta_r=0.07$ and $\Delta_{\theta}=0.17$ for Run~C3. 
Hence, our model is even closer to the anti--solar
to solar--like differential rotation transition than Run~C3 
of theirs.
Therefore, the contribution of the differential rotation to the 
large-scale dynamo should be negligible.
To assert this, though, a more thorough analysis 
would be required.

In comparison to Run~D in \cite{VWKKOCLB18} (Figure~\ref{fig:KrComp}, 
second panel), which had 
weak values for the relative differential rotation 
($\Delta_r=0.003$ and $\Delta_{\theta}=0.007$), Run~R2 presents stronger 
DR in absolute value, although with the opposite sign.
The rotation profile is anti--solar with a retrograde flow at the equator.
At mid--latitudes, a region of accelerated flow develops and the 
isorotation contours here and at higher latitudes are 
radially inclined.
In the thick RZ the rotation does not vary much in latitude
and depth.

The rotation profile of Run~R3 (\Figu{fig:KrpOm}) 
is very similar to the one from its wedge counterpart shown in
Figure~\ref{fig:KrComp}, third panel. It is solar--like,
showing an accelerated equator, and 
has a rather weak relative differential 
rotation in terms of $\Delta_r$ and $\Delta_{\theta}$.
The minimum at mid--latitudes is present, and its location 
corresponds to the sub-adiabatic region at the top boundary,
which is probably numerical in nature.
A near--surface shear layer, with a negative radial gradient,
is present from mid to low latitudes. 

The rotation rate of Run~R4 corresponds to Run~H in 
\cite{VWKKOCLB18}, while its Coriolis number is close
to Run~G$^{a}$ in the same study,
see also the last two panels in Figure~\ref{fig:KrComp}.
Hence, the usage of the Kramers opacity law 
produces higher convective velocities and, therefore, smaller $\Co$.
The values for $\Delta_r$ and $\Delta_{\theta}$ coincide with the
ones from G$^{a}$.
The rotation profile, shown in \Figu{fig:KrpOm}, rightmost panel,
closely resembles that of Run~H, with a deep minimum at 
mid--latitudes.

\subsection{Dynamo solutions}\label{sec:magfield}
\begin{table*}[h]
    \centering
    \begin{tabular}{c c c c c c c c c c}
    \hline
      Run & E$_{\rm mag,tot}^{\rm dec}$ & E$_{0}^{\rm dec}$ & E$_{1}^{\rm dec}$ & E$_{2}^{\rm dec}$ & E$_{3}^{\rm dec}$ & E$_{4}^{\rm dec}$ & E$_{5}^{\rm dec}$ & E$_{l, m > 5}^{\rm dec}$ & $\tau_{\rm cyc}$[yr] \\
      \hline\hline
      R1 &1.4(-2) & 4.9(-3) & 1.6(-3) & 8.2(-4) & 7.5(-4) & 8.0(-4) & 8.5(-4) & 4.5(-3) & $1.8_{\rm (m=0)}$\\
      R2 & 1.5(-2) & 9.9(-3) & 3.7(-4) & 3.9(-4) & 4.2(-4) & 4.1(-4) & 4.8(-4) & 2.9(-3) & $1.7_{\rm (m=0)}$\\
      R3 & 1.8(-2) & 5.4(-3) & 8.2(-3) & 1.2(-3) &    7.5(-4) & 5.2(-4) & 4.4(-4) & 1.0(-3) & $3.3_{\rm (m=0)}$ \\
      R4 & 5.0(-2) & 1.2(-2) & 2.6(-2) & 3.4(-3) &     1.7(-3) & 1.2(-3) & 1.0(-3) & 4.2(-3) & $2.8_{\rm (m=1)}$ \\
    \hline
    \end{tabular}
    \caption{Magnetic energy from the decomposition in the first 
    11 spherical harmonics ($0\leq l,m\leq 10$) of the near-surface
    ($r=0.98R$) radial magnetic field.
    The labels E$_{m}^{\rm dec}$ indicate the energy in the
    corresponding $m$ mode, in units of $10^{5}~{\rm J~m^{-3}}$.
    We define $0\leq l,m \leq 5$ the large--scale field.
    The numbers in parenthesis represent the power of ten.
    Last column indicates the characteristic time of the dynamo,
    calculated on the dominant mode,
    indicated in the subscript. 
    }
    \label{tab:Krdec}
\end{table*}

All the presented runs develop large--scale dynamo 
(LSD) action, and thereby sustain a magnetic field. 
The magnetic Reynolds numbers, however, are too low
to allow for small--scale dynamo action (SSD). 
Our dynamo solutions, therefore, exhibit magnetic fields 
on the largest scales, but also a strong fluctuating
component, that is generated by tangling of the 
LSD--generated magnetic field by the turbulent motions 
rather than from an SSD.
We present the results of the decomposition 
of the magnetic field
in the first 11 spherical
harmonics ($0\leq l,m \leq 10$) in Table~\ref{tab:Krdec}.
The decomposition was performed on the radial magnetic field component
near the surface of the simulation ($r=0.98R$).
For each of the runs, we calculate the characteristic time of variation
of the magnetic field, $\tau_{\rm cyc}$, 
from the time evolution of the 
dominant dynamo mode.
The results are shown in Table~\ref{tab:Krdec}, last column, and
the mode from which $\tau_{\rm cyc}$ is calculated is indicated as a 
subscript.
As described in \cite{VWKKOCLB18}, these cycles
are at most quasi--periodic, hence Fourier analysis
is not suitable here. Instead, we use a syntactic
method, which means that we 
count how many times the dominant mode of 
the magnetic field peaks above its mean value,
and $\tau_{\rm cyc}$ is obtained by dividing
the length of the full time span of measurement by
the number of peak times.

Run~R1 and R2 have a dominant axisymmetric large--scale magnetic field, 
but also a significant contribution from the small--scale field 
($l,m >5$).
The energy in the first non--axisymmetric large--scale mode is 
less than in the mode $m=0$.
This is opposite to the case in \cite{VWKKOCLB18}, where simulations 
with the same rotation rates, but 
a fixed heat conduction profile, were showing a substantial $m=1$
component.
Run~R3 is in a regime where the axisymmetric and the 
first non--axisymmetric mode have comparable strengths,
hence we characterize this run as being non-axisymmetric.
R3, in fact, shows a weak azimuthal dynamo wave (ADW, see, also 
\ref{subsub:naxi}).
For calculating $\tau_{\rm cyc}$, however, we follow the same convention as
in \cite{VWKKOCLB18} for runs in this regime, and use $m=0$ to obtain it.
Run~R4 has a strong $m=1$ component, which is reflected by the 
presence of the ADW in Figure~\ref{fig:KrADW}, lower panel.

\subsubsection{Axisymmetric magnetic field}\label{subsub:axi}

\begin{figure*}
\centering
\includegraphics[width=0.48\textwidth]{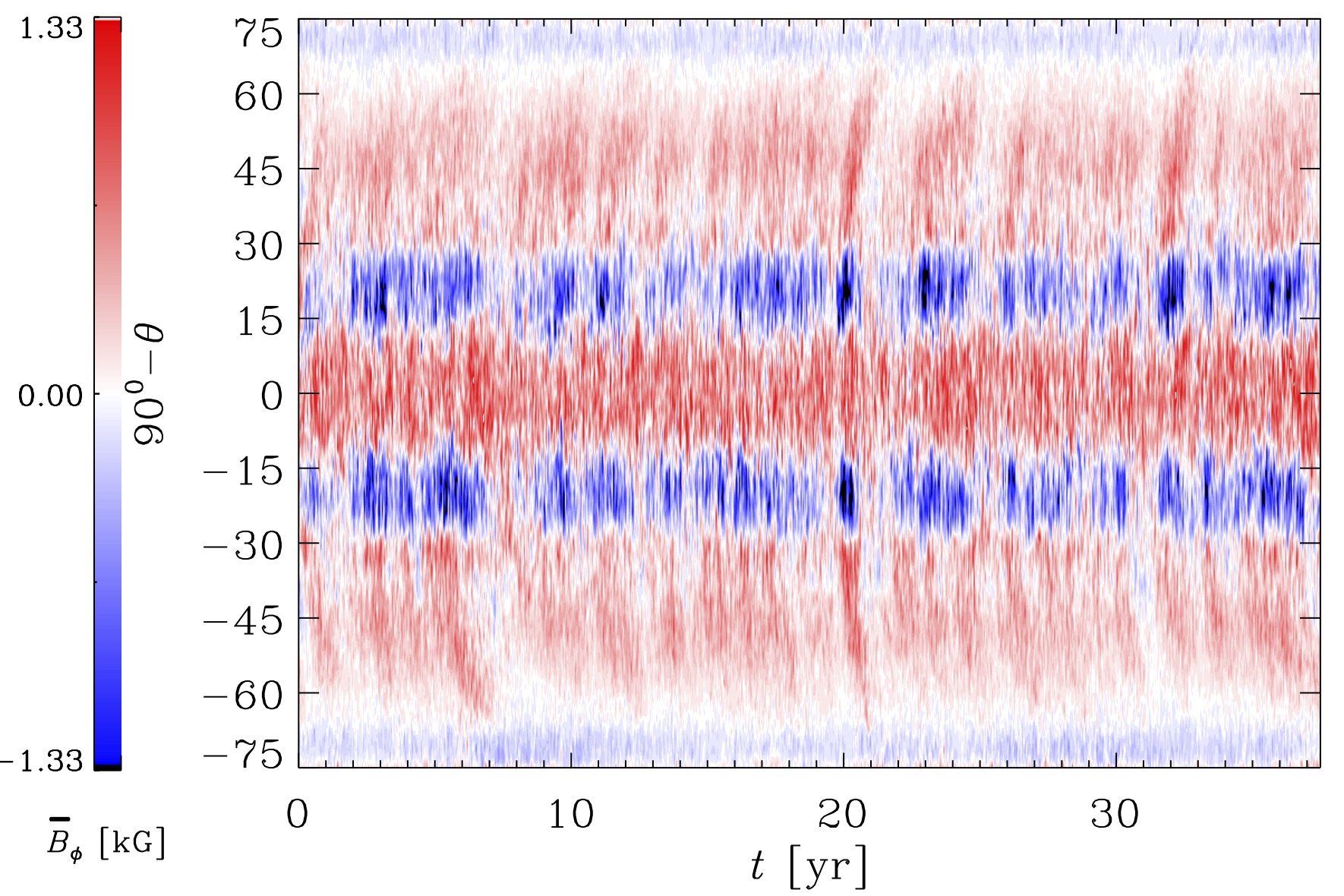}
\includegraphics[width=0.48\textwidth]{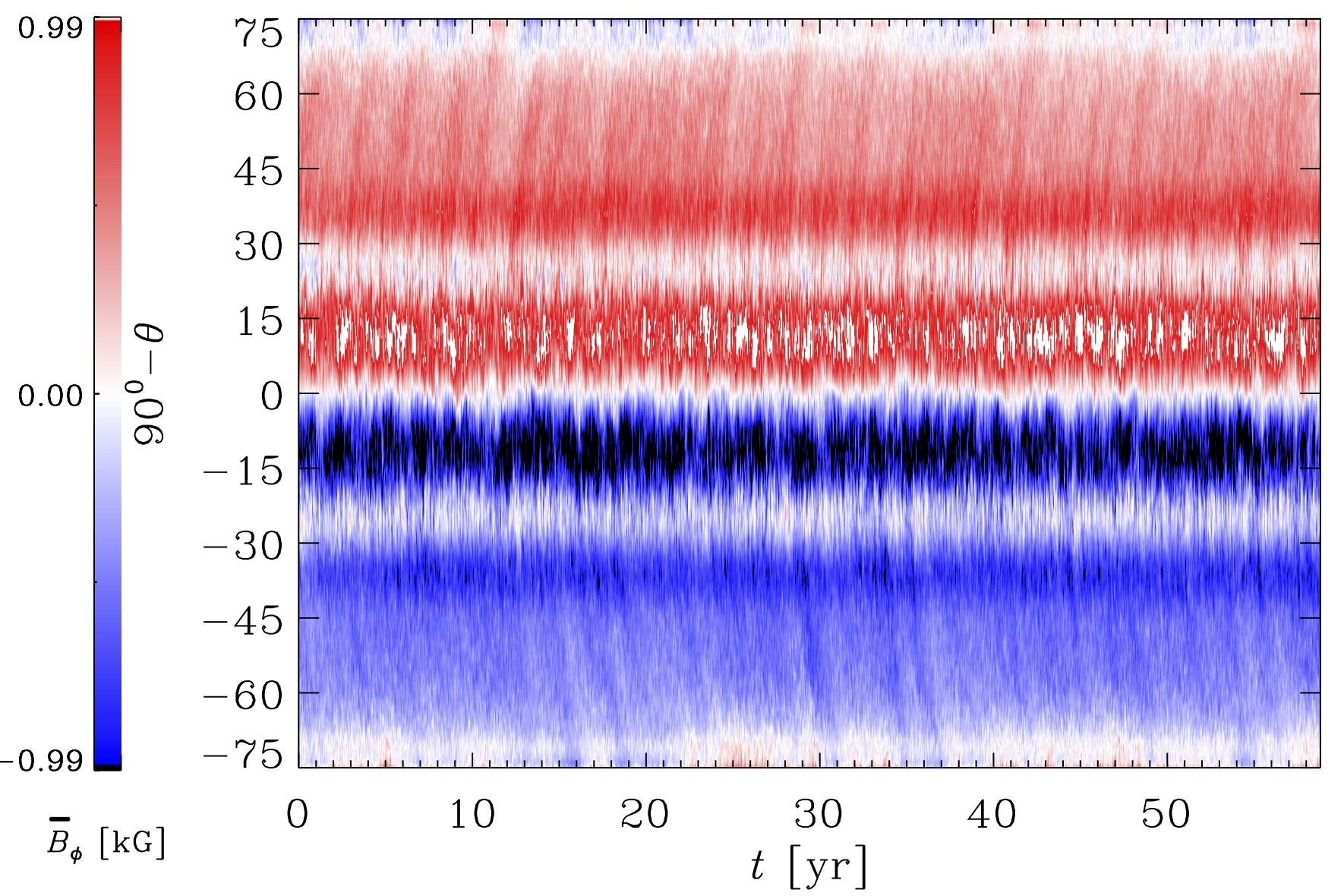}
\includegraphics[width=0.48\textwidth]{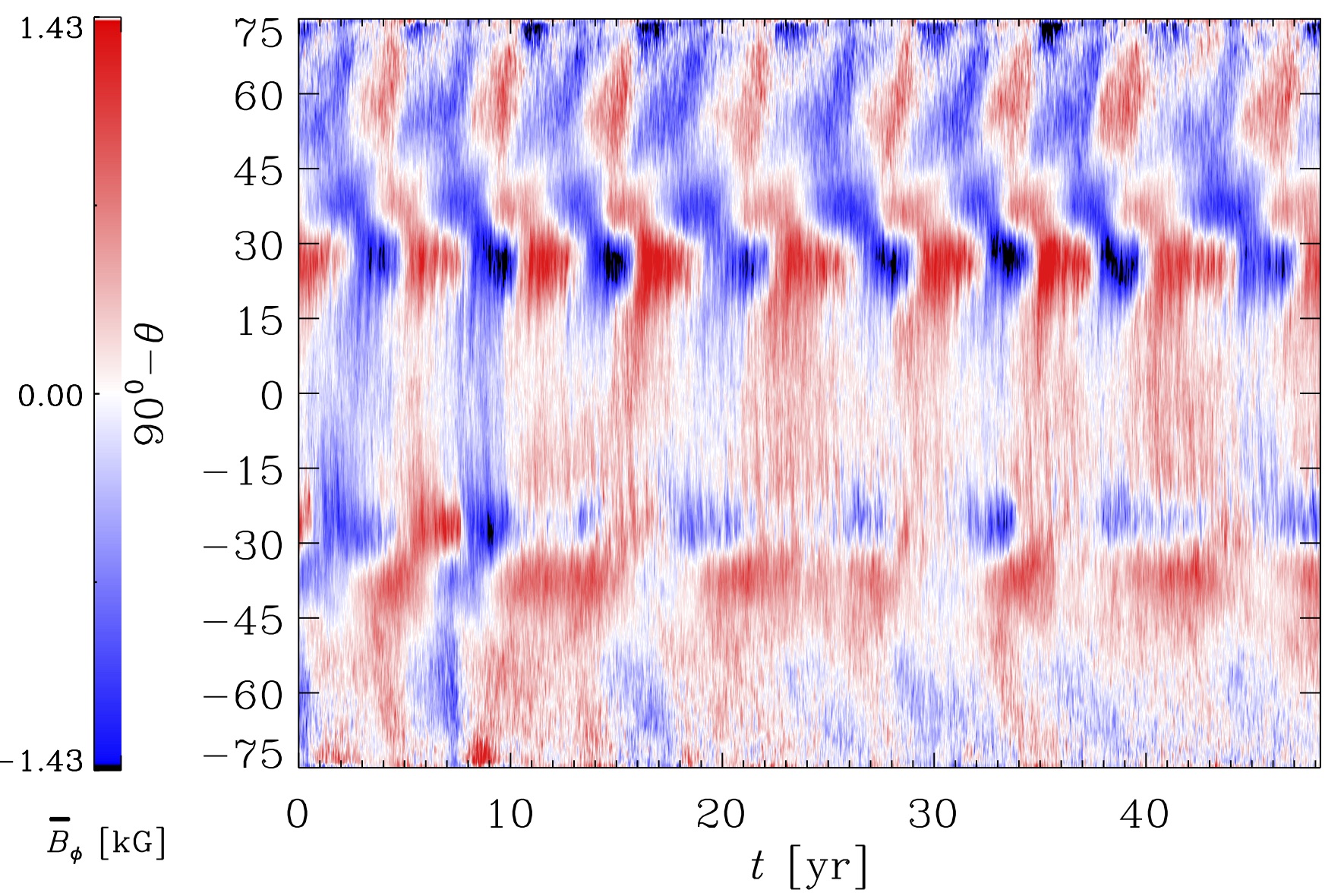}
\caption{Azimuthal magnetic field at $r=0.98R$ for Run~R1 and Run~R3.}\label{fig:Krpbut}
\end{figure*}

We show the azimuthally
averaged longitudinal magnetic field near the surface
as a function of time (the so-called butterfly diagram) in
Figure~\ref{fig:Krpbut}. 
Run~R1 is characterized by equatorially symmetric
magnetic field, with non--migrating negative polarities at
low latitudes and 
poleward migrating positive field 
at higher latitudes.
A stationary negative field is present at all times close to
the latitudinal boundary.
A similar, oscillatory dynamo solution
was reported and analysed in detail
in \cite{VKWKM19}. There it was concluded that
two dynamo modes were competing in the model,
a stationary and an oscillatory one, the latter with polarity reversals.
This dynamo was concluded to be
driven mostly by turbulent effects, as the differential rotation was 
found weak in the model.
Run~R1 appears to be another incarnation of such a dynamo in the transition
regime from solar--like to anti--solar differential rotation.

\begin{figure}
\centering
\includegraphics[width=\columnwidth]{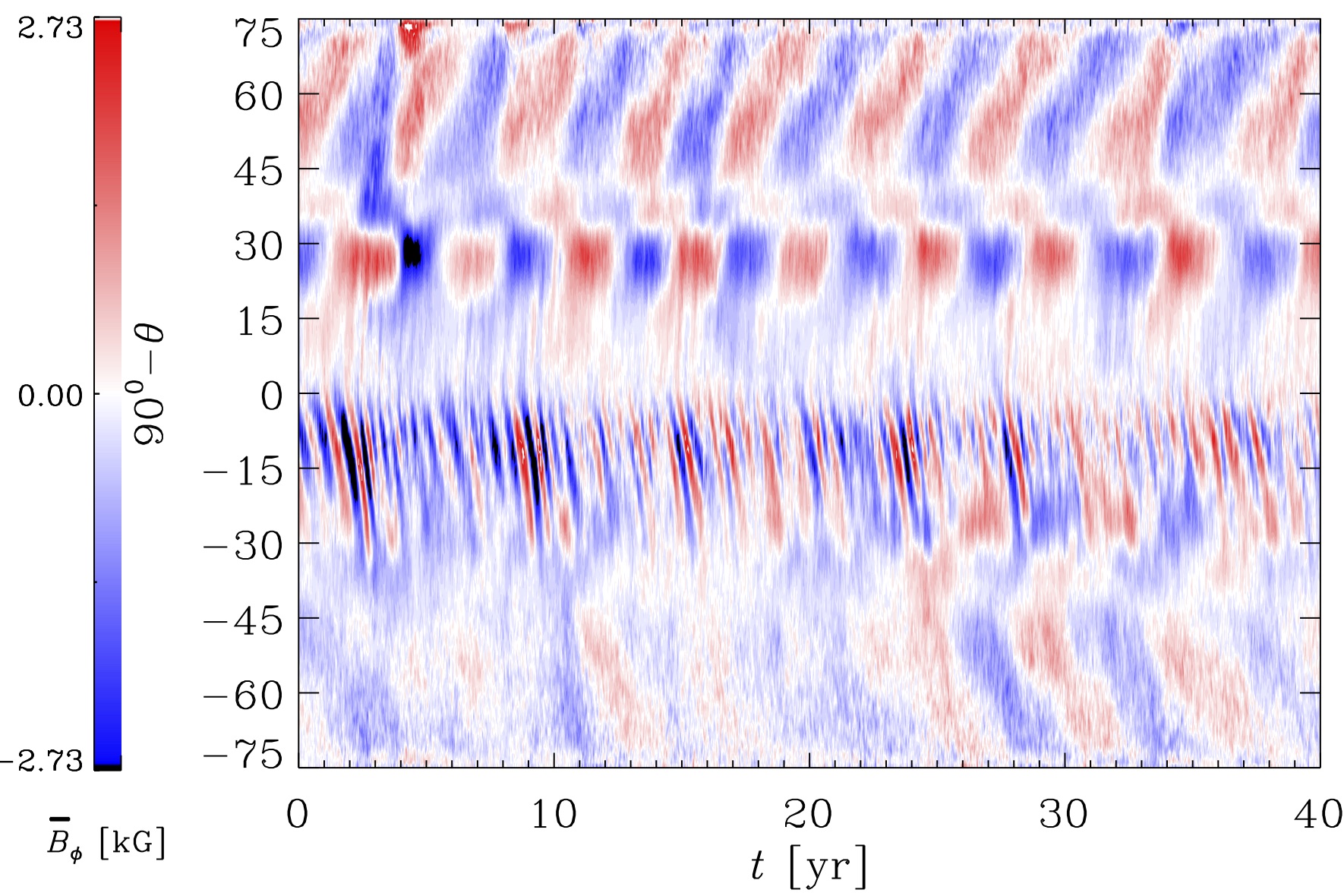}
\includegraphics[width=\columnwidth]{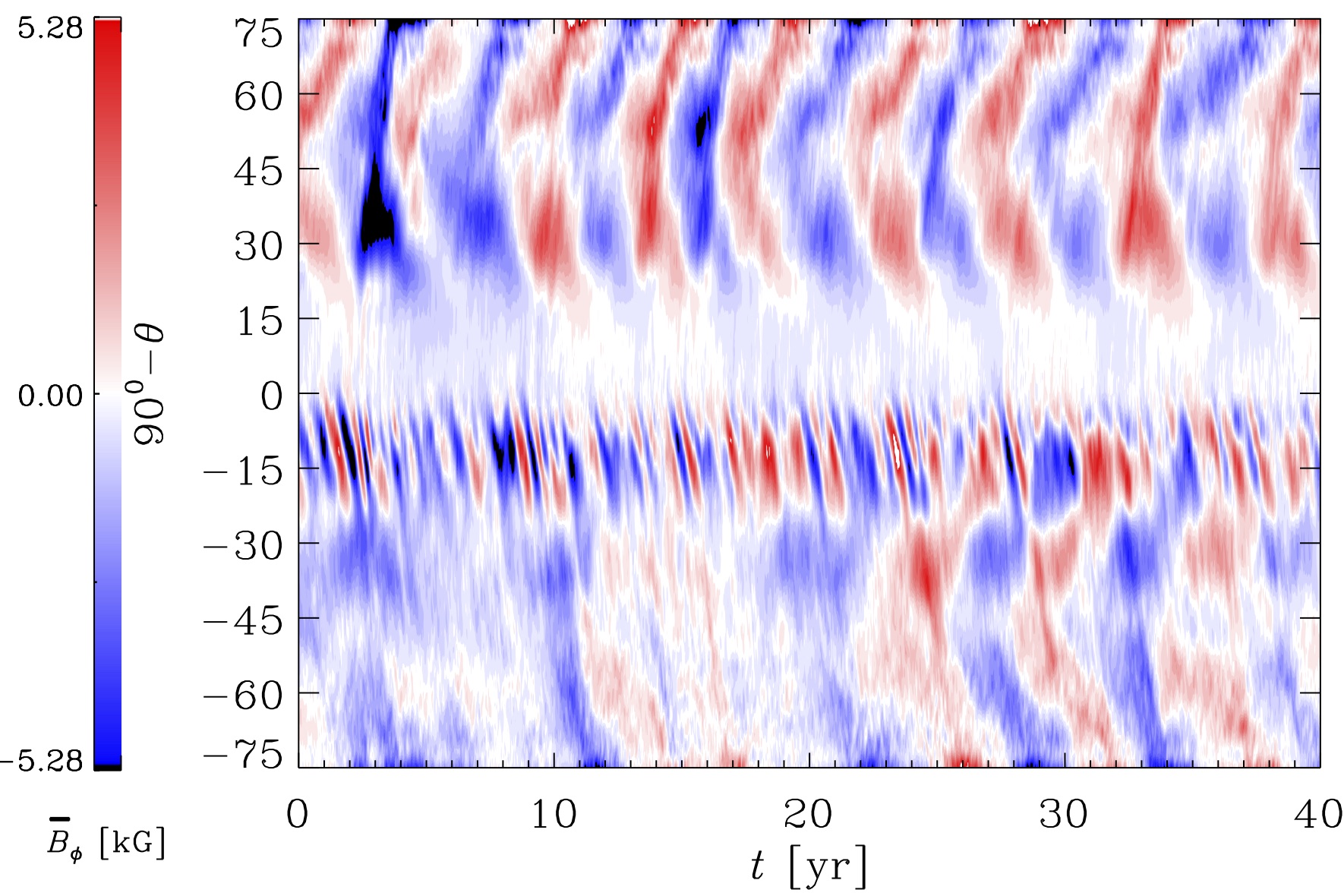}
\includegraphics[width=\columnwidth]{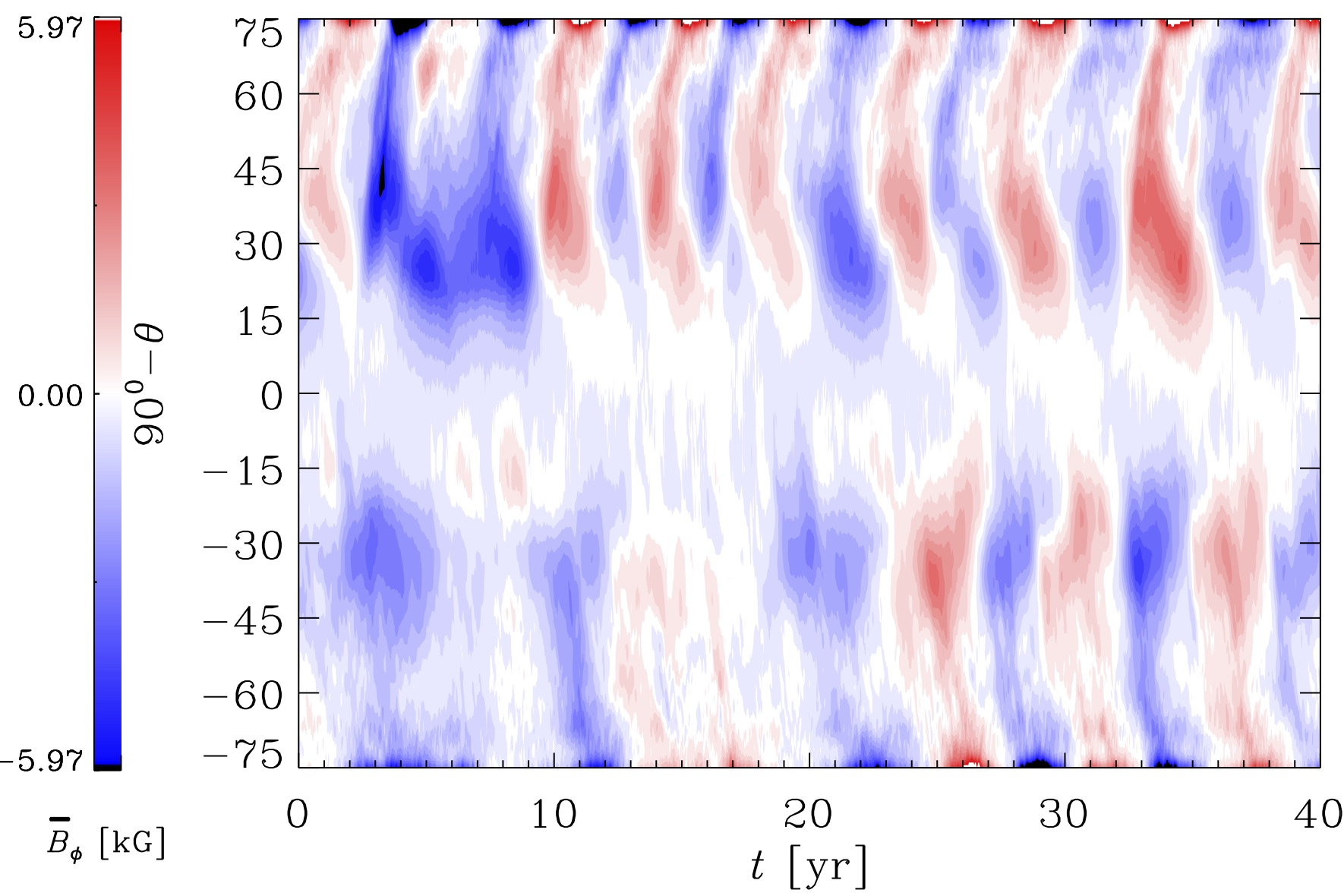}
\caption{Azimuthal magnetic field for Run~R4 at different depths:
$r=0.98R$, $r=0.86R$ and $r=0.75R$.}\label{fig:Krpbut6Om}
\end{figure}

The longitudinally averaged $B_{\phi}$ in Run~R2 is dipolar 
(equatorially antisymmetric) at the surface.
The polarity is positive in the 
upper
hemisphere and there
are no signs of polarity reversals, 
if the weak ones at the latitudinal 
boundaries are not counted for. 
Also, no equatorward migration can be seen, albeit there is again a
tendency for a weak poleward migration with a very high frequency.

Run~R3 exhibits equatorial propagation of the 
azimuthal magnetic field, and the pattern in the 
butterfly diagram (lowest panel in Fig.~\ref{fig:Krpbut})
is very similar to that of Run~MHD2 in \cite{KVKBF19}, 
also showing a similar periodicity of $\sim 2~{\rm yr}$.
In contrast to Run~MHD2, the solution shows a pronounced hemispheric
asymmetry, with a regular cycle in the upper hemisphere and 
an irregular periodicity in the lower one.
The latter cycle seems to be longer than the one in the 
upper hemisphere.

In Fig.~\ref{fig:Krpbut6Om} we show butterfly diagrams at 
three different depths for Run~R4.
At the surface, it shows two dynamo modes: a high--frequency 
one in the lower hemisphere and a lower--frequency one,
with a periodicity similar to Run~R3, in the upper hemisphere.
As we go deeper down to the convection zone, the high--frequency 
mode disappears, and we can trace
its origin to the depth of $0.80 \leq r\leq 0.85$, therefore to 
the bottom of the BZ.
The lower--frequency mode, however, persists until
the OZ, and therefore we infer that it is generated
there.
The existence of different dynamo modes at different depths has been 
reported already in other studies 
\citep[such as][]{KKOBWKP16,KVKBF19}.

\subsubsection{Non-axisymmetric magnetic field}\label{subsub:naxi}
\begin{figure}
\centering
\includegraphics[width=\columnwidth]{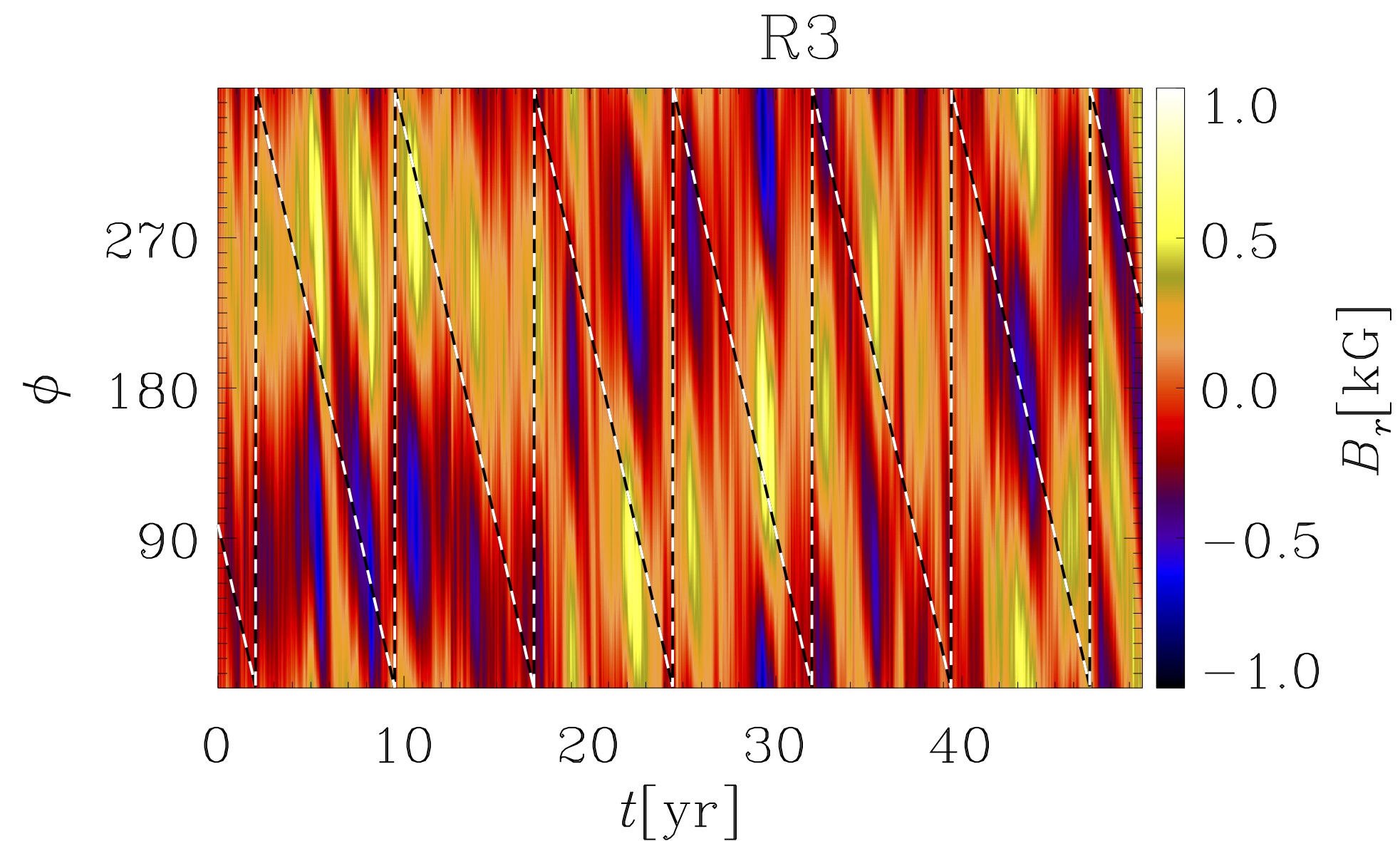} 
\includegraphics[width=\columnwidth]{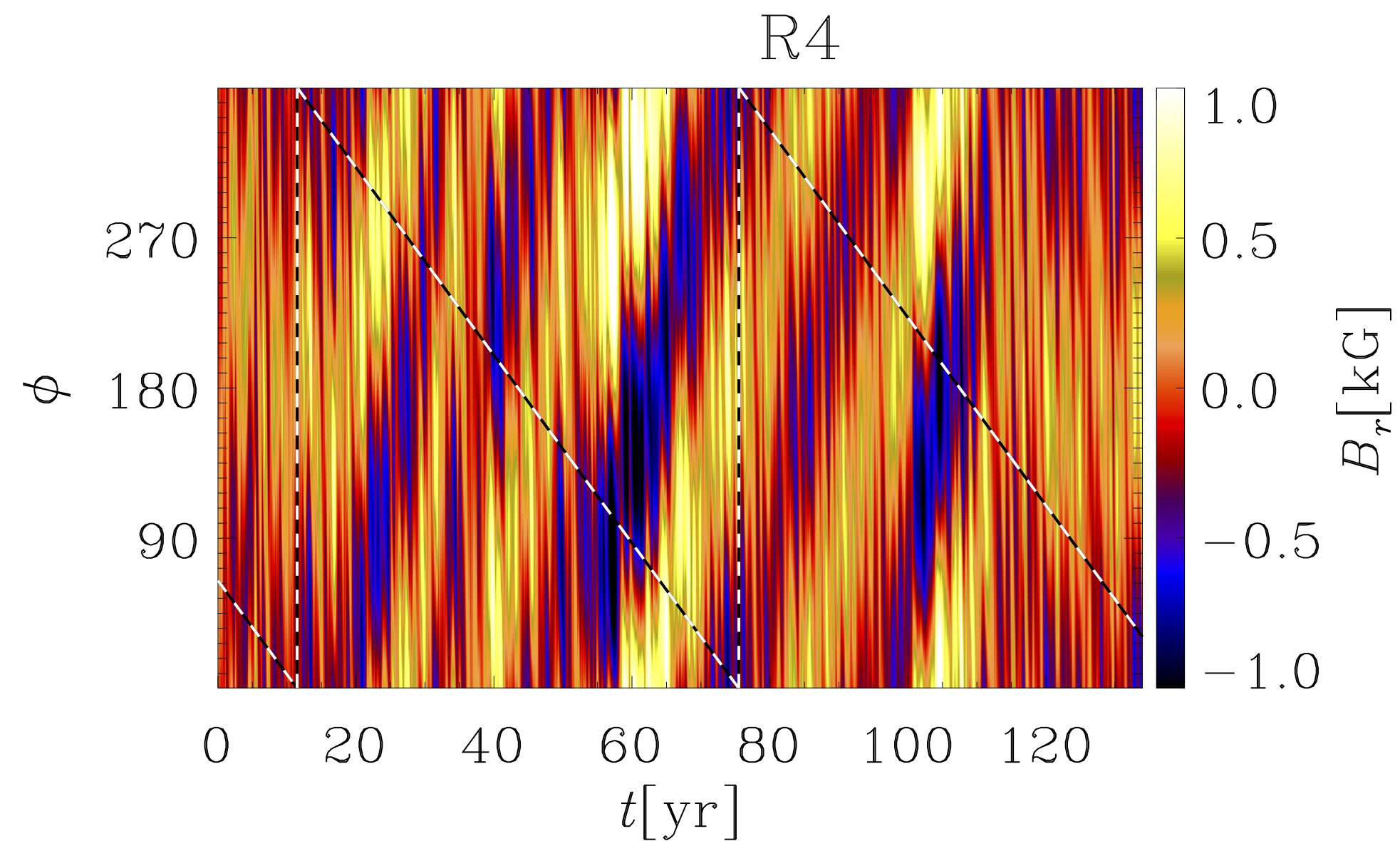}
\caption{Azimuthal dynamo wave for Runs~R3 and R4 as a function of longitude and time, at latitude
$\theta=+45^o$ and depth $r=0.98R$.
The black and white dashed line represents the differential
rotation at the same latitude.}\label{fig:KrADW}
\end{figure}

In Run~R3 a weak azimuthal dynamo wave is present.
In Figure~\ref{fig:KrADW}, upper panel, 
we plot the reconstructed $m=1$ mode at $45^o$ above the equator
close to the surface as a function of time and longitude.
The black--white dashed line represents the pattern of 
differential rotation at the same latitude.
In the absence of a dynamo wave, the magnetic field would follow 
the propagation speed of this pattern.
Instead, in \Figu{fig:KrADW}, the magnetic field does not fall 
on the line for most of the time, hence, it has its own motion 
as a wave, travelling in the prograde direction.
Weak ADWs, such as the one seen in the case of Run~R3, 
were found to be typical in simulations that are close to the 
axi- to non-axisymmetric transition \citep{VWKKOCLB18}.
In this study it was observed that,
when the energy in the modes $m=0$ and $m=1$ is comparable,
the ADW can be affected by the differential rotation,
in which case it becomes advected by it
for some time intervals (see \Figu{fig:KrADW}, left panel, 
$0~{\rm yr} \leq t\leq 15~{\rm yr}$).
ADWs were already found in other numerical studies
\citep{KMCWB13,CKMB14,VWKKOCLB18}, but their direction was mostly
retrograde, in contrast with observational results 
\citep[see, e.g.,][]{Lehtinen16}.

A stronger prograde ADW is also present in Run~R4.
The wave does not persist at all times, but there are periods 
when it disappears.
Similar behaviour was also observed, for example, in the 
temperature maps of the active star II Peg in \cite{Marjaana11},
where a very clearly defined prograde ADW persisted over 
ten years, but then vanished. 
Also \cite{2012A&A...542A..38L} and \cite{Olspert15} reported on 
ADWs on the young solar analogue star LQ Hya, but these lasted 
even 
for
a shorter period of time,
of about a couple of years. 
We attribute the change in the ADW direction to the different
heat conduction prescription in these runs.
Moreover, in Run~R3 the ADW can even change direction from prograde 
to retrograde during some short epochs, most likely related to the stronger
influence from the differential rotation. Such a change of direction in the
migration of active longitudes has also been observed, e.g. in the study of
\cite{Korhonen04} in the case of the intensively studied single active star 
FK Coma Berenices.

We calculate the period of the ADW, $P_{\rm ADW}$ as in
\cite{VWKKOCLB18}, taking the latitudinal and temporal average
of the slope of the reconstructed time evolution of the $m=1$ mode.
For R3, we obtain $P_{\rm ADW}=-24.5~{\rm yr}$, the minus sign
indicating a negative slope, therefore a retrograde direction for 
the dynamo wave, in contrast with 
the mostly prograde appearance of the migration in
Figure~\ref{fig:KrADW},
upper panel.
As discussed before, the ADW in R3 is chaotic,
even changing its direction, and 
also disappearing,
and this causes the average period to be an inaccurate measure.
$P_{\rm ADW}=44.9~{\rm yr}$ for Run~R4, the positive sign indicating 
a prograde wave, as expected.
From the lower panel of Figure~\ref{fig:KrADW}, however, 
we would infer a shorter period $\sim~30~{\rm yr}$.

\section{Conclusions}\label{sec:KrConcl}
In this paper we studied the effect of a dynamically adapting 
heat conduction prescription, based on Kramers opacity law, on
semi--global MHD simulations.
The main aim was to determine its effect on the two major transitions
reported in numerical studies \citep[e.g.,][]{GYMRW14,VWKKOCLB18}. 
One concerns the rotation profiles, and is the transition from
accelerated poles and decelerated equator
to a solar--like profile, with faster equator.
The other one involves the large--scale magnetic field, and is the
transition from an axisymmetric magnetic field, as in the Sun,
to a non--axisymmetric one found in more rapid rotators.
Previous studies \citep[][]{VWKKOCLB18} reported these transitions
occurring at the same rotation rate, in contrast with 
the current interpretation of observations.
The fact that simulations usually produce anti--solar 
differential rotation for the solar rotation rate could indicate
that the Sun is in a transitional regime 
\citep[e.g.,][]{KKB14, 2016ApJ...826L...2M}, or it could also mean that
simulations still cannot fully capture the right rotational
influence on turbulent convection in the Sun.
The study of \cite{Lehtinen16} reported on 
the existence of non--axisymmetric structures in stars with
varying rotation rates, and hence could 
determine quite a sharp transition point, in terms of 
the rotation period, when fields turn from axi-- to 
non--axisymmetric configurations. 
According to dynamo theory, these two modes can compete, 
and there can be a transition region, where both dynamo 
modes co--exist, as is also clearly demonstrated by the models 
presented in this paper and \cite{VWKKOCLB18}. 
Hence, the observational transition point must be regarded 
as a lower limit, in terms of the rotation period, for the 
transition, as it could be that the sensitivity of the current 
instruments is not high enough to detect the very weak 
non--axisymmetric components.
Since active longitudes have not been detected 
on the Sun \citep{PBKT06}, though,
these two transitions should not be located at 
the same, nearly solar, rotation rate.

In runs with slow rotation, the 
differential rotation profile is significantly 
affected by the Kramers opacity law and, as a result, solutions 
with less solar--like characteristics, like an almost rigid body
rotation and a minimum at mid--latitudes, develop.
The different heat conduction prescription also promotes 
the formation of a stably stratified layer, rather isotropic in
latitude, in the lower quarter of the domain.
For faster rotating runs, the rotation profile is solar--like,
but still maintains the minimum at mid--latitudes, and a latitudinally
changing subadiabatic region forms near the equator.
Also, the Coriolis number is lower than in the corresponding cases
using fixed profiles for heat conduction,
which is most likely the largest contributing factor to push 
the anti--solar to solar differential rotation transition to 
an unwanted direction of more rapid rotation rates.

The convective transport is efficient, isotropic and almost
radial everywhere in the convective region 
in models with slow rotation (Runs~R1 and R2), while it gets strongly
concentrated to the equatorial region in runs with more rapid rotation
(Runs~R3 and R4).
Also, the BZ becomes shallower close to the tangent cylinder 
in the rapid rotation regime. 
Moreover, hemispheric asymmetries in the convection
zone structure are seen in the run with the fastest rotation (Run~R4).

The large--scale magnetic field is axisymmetric in Run~R1
and R2, while for Run~R3 and Run~R4 the first
non--axisymetric mode is dynamically more important.
Both the fast rotating runs have a hemispherically asymmetric
oscillating magnetic field, with a periodicity of $\sim~2$ years.
As in \cite{VWKKOCLB18}, the magnetic cycle lengths do not 
overall depend strongly on the rotation period.
The strong magnetic field in all the runs originates
from the subadiabatic layer.
In Run~R4 a high--frequency mode is present in the 
lower hemisphere.
This component is generated at the bottom boundary of the BZ.
The co--existence of multiple dynamo modes at different depths of the 
convection zone is consistent with previous studies
\citep[e.g.,][]{KKOBWKP16}, using prescribed profiles for
heat conduction.
In this study, the high--frequency mode was generated 
near the surface, while the low--frequency one in the middle 
of the CZ.

In the non--axisymmetric runs, ADWs are present: a weak one for 
Run~R3 and a stronger one for Run~R4.
In both cases, the direction is prograde, in agreement with 
photometric observations \citep[][]{Lehtinen16}.
In the previous numerical study using a 
prescribed heat conduction profile \citep{VWKKOCLB18}, we found 
a preference for retrograde ADWs.
The ADWs also show time variations.
For Run~R3, the ADW is rather weak and the differential rotation
can advect it for some time, changing the direction of the wave.
This could be caused by the comparable relative energies
in the $m=0$ and $m=1$ modes.
In Run~R4 the stronger ADW disappears at certain times.
Such behavior is also what is observed for active stars
\citep[e.g.,][]{Korhonen04,Marjaana11,2012A&A...542A..38L,Olspert15}, 
where the active longitudes disappear
or have the same velocity as the surface rotation.

In summary, in this study we have shown that both the
major transitions related to stellar dynamos are affected by the use of 
a more physical description of heat conduction in 
global magneto--convection simulations.
The differential rotation profiles undergo a significant change
near the anti--solar to solar--like
differential rotation transition, but all the runs are still 
in Taylor--Proudman balance, with almost cylindrical isocontours.
For the same rotation rates, the convective velocities are higher, 
hence Coriolis numbers are lowered,
resulting in an anticipated transition to anti--solar differential
rotation, in contrast with observations.
The transition from axi-- to non--axisymmetric magnetic fields is shifted 
towards higher rotation rates. 
The direction of the ADW is reverted with respect to previous 
studies, producing a better agreement with observations.

\begin{acknowledgement}
MV and MJK acknowledge support from the Independent Max Planck Research Group "SOLSTAR". 
MV acknowledges the support from the International Max Planck Research School 
for Solar System Science at the University of G\"ottingen.
MJK acknowledges the support of 
the Academy of Finland ReSoLVE Centre of Excellence (grant No.~307411).
This project has received funding from the European Research Council
under the European Union's Horizon 2020 research and innovation 
programme (project "UniSDyn", grant agreement n:o 818665).
The authors would like to thank Petri K\"apyl\"a, J\"orn Warnecke
and Matthias Rheinhardt for constructive comments and conversations.
\end{acknowledgement}

\appendix
\section{Comparison with rotation profiles in previous
works}\label{sec:appendix}

\begin{figure*}
\centering
\includegraphics[width=0.19\textwidth]{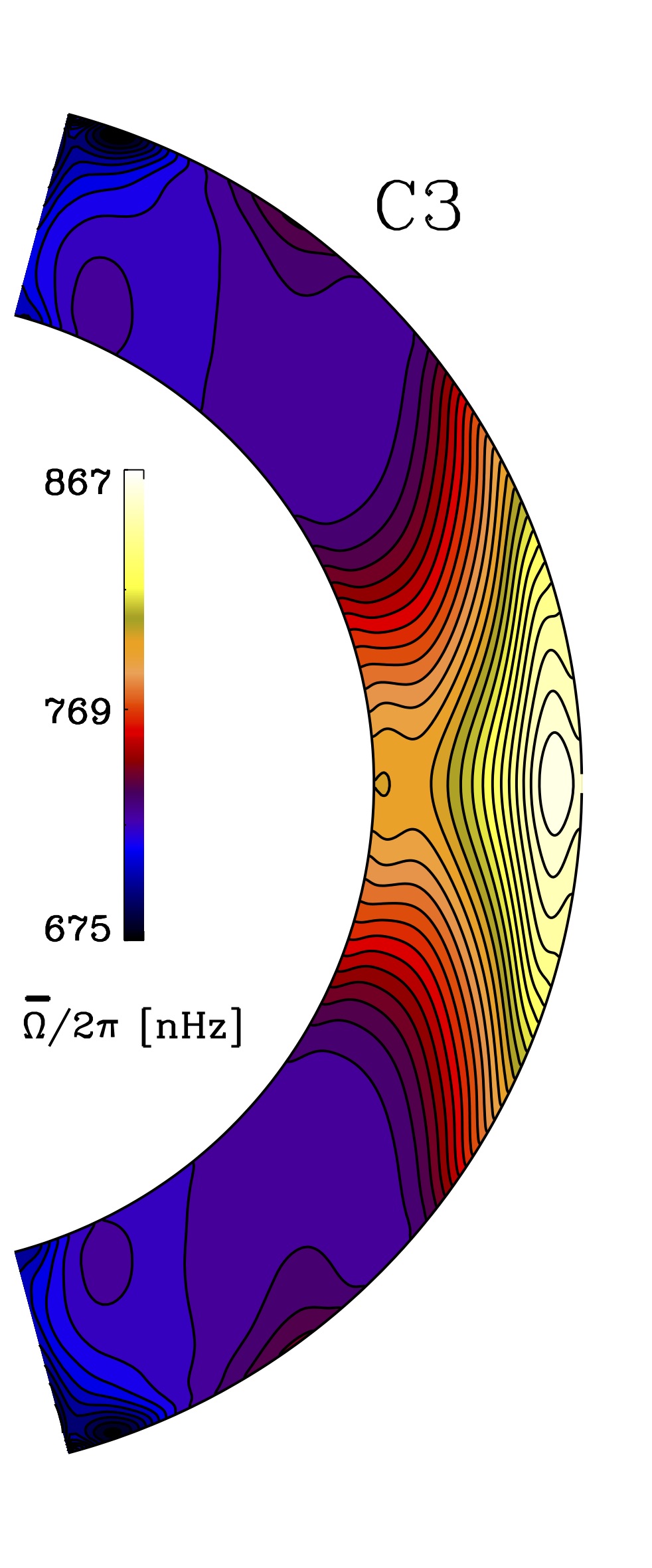} 
\includegraphics[width=0.19\textwidth]{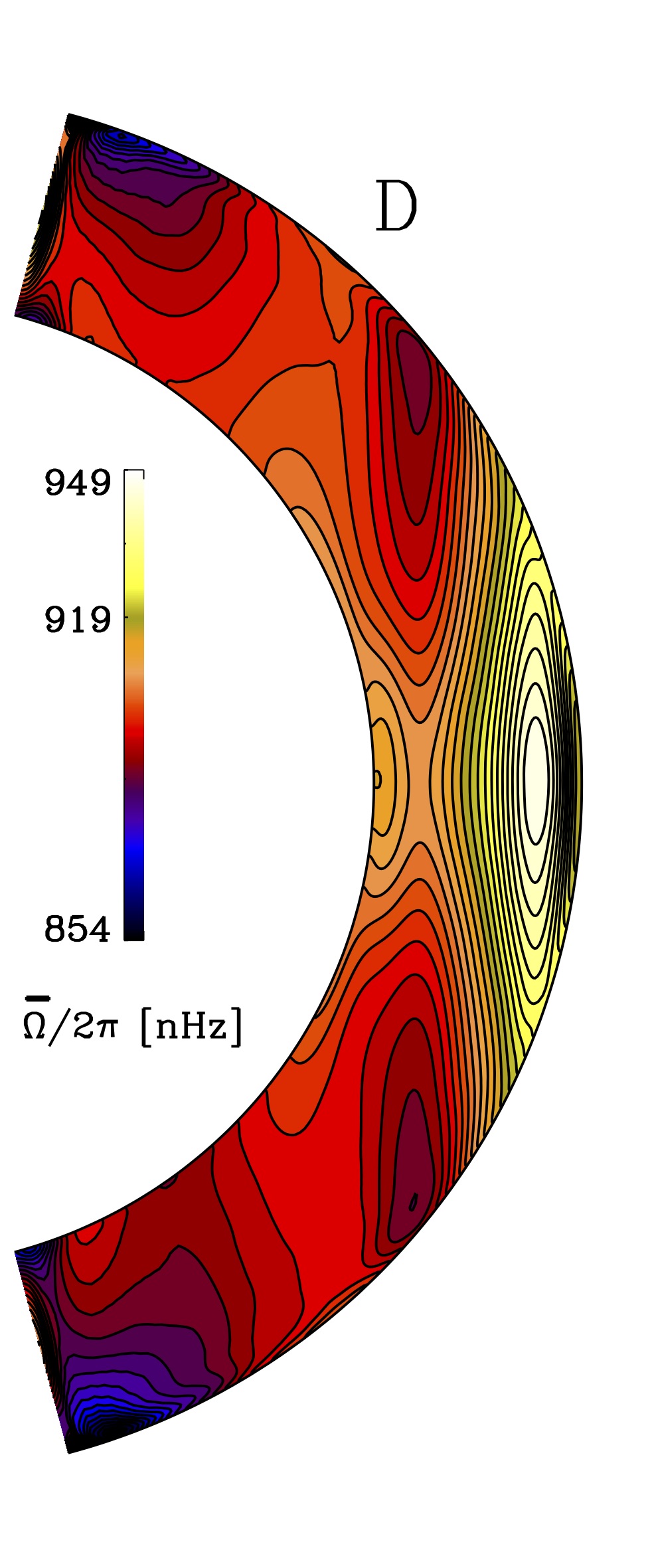}
\includegraphics[width=0.19\textwidth]{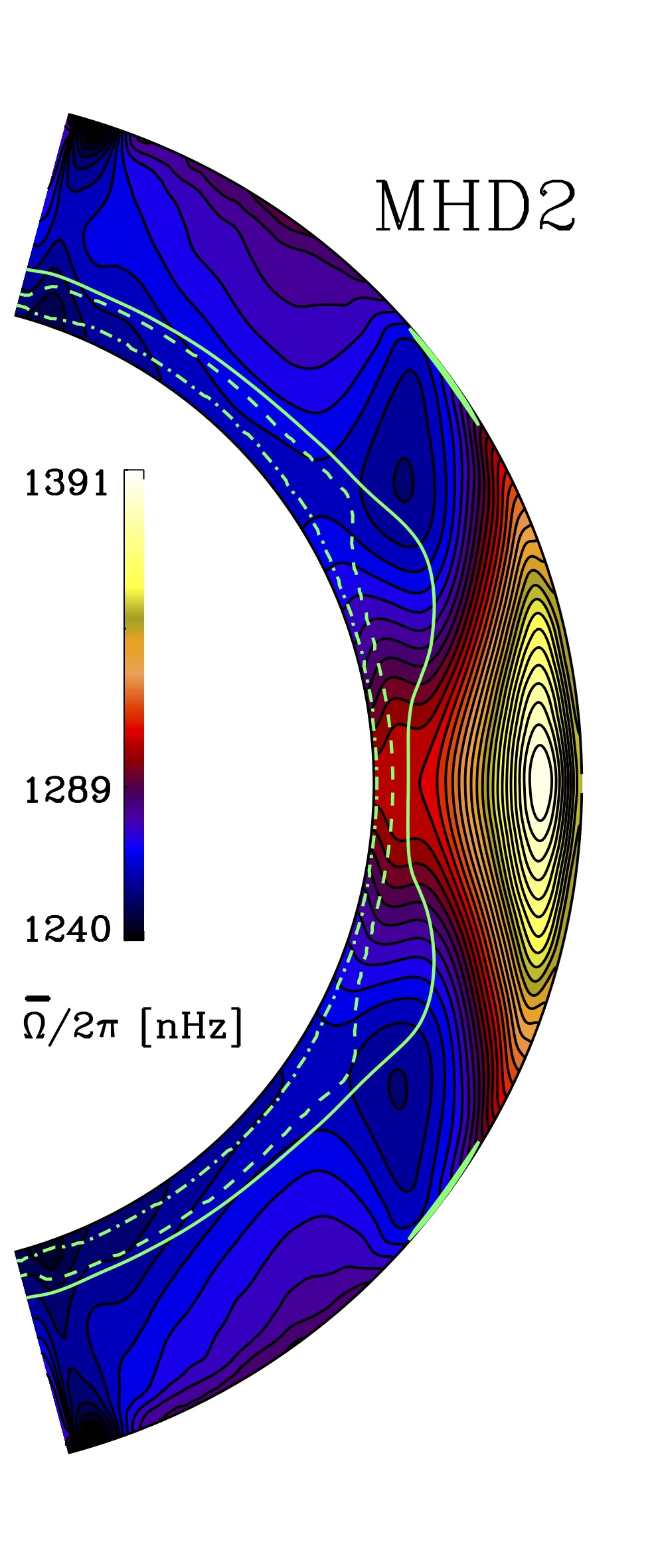}
\includegraphics[width=0.19\textwidth]{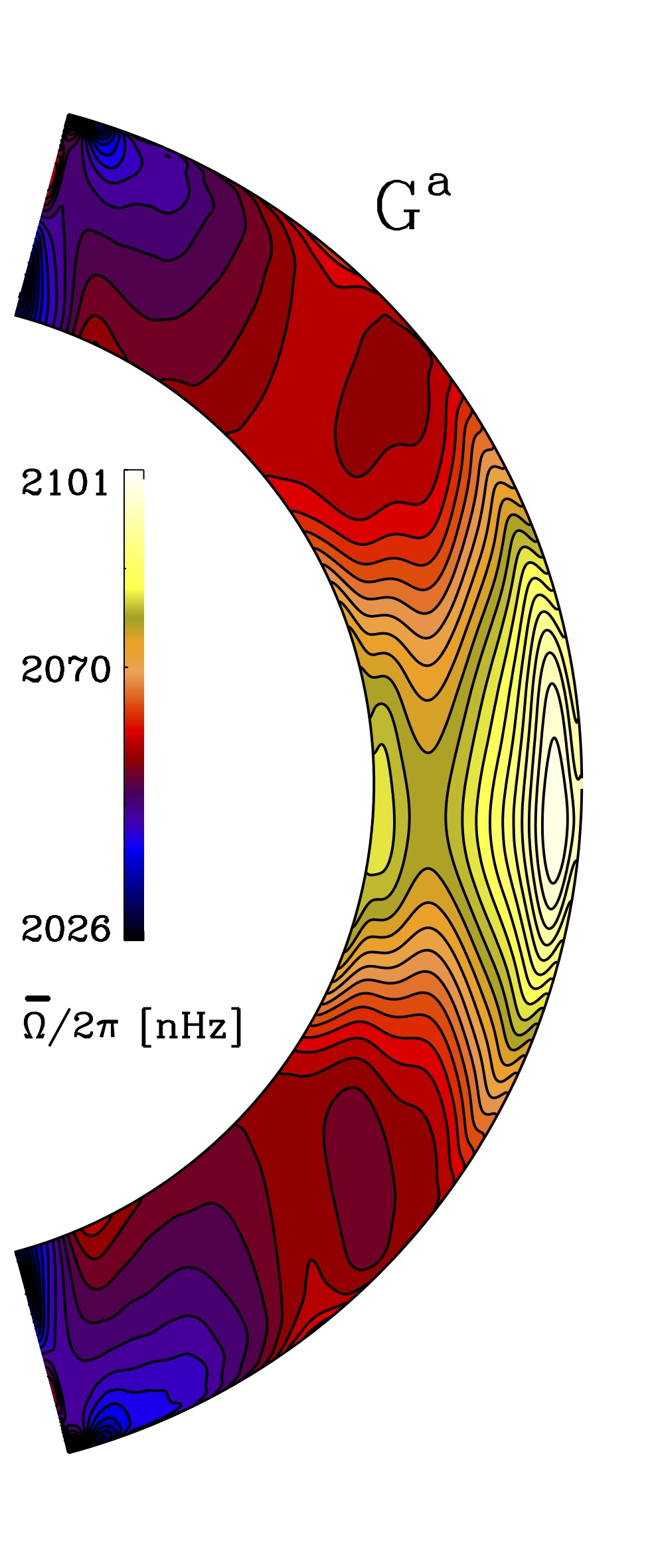}
\includegraphics[width=0.19\textwidth]{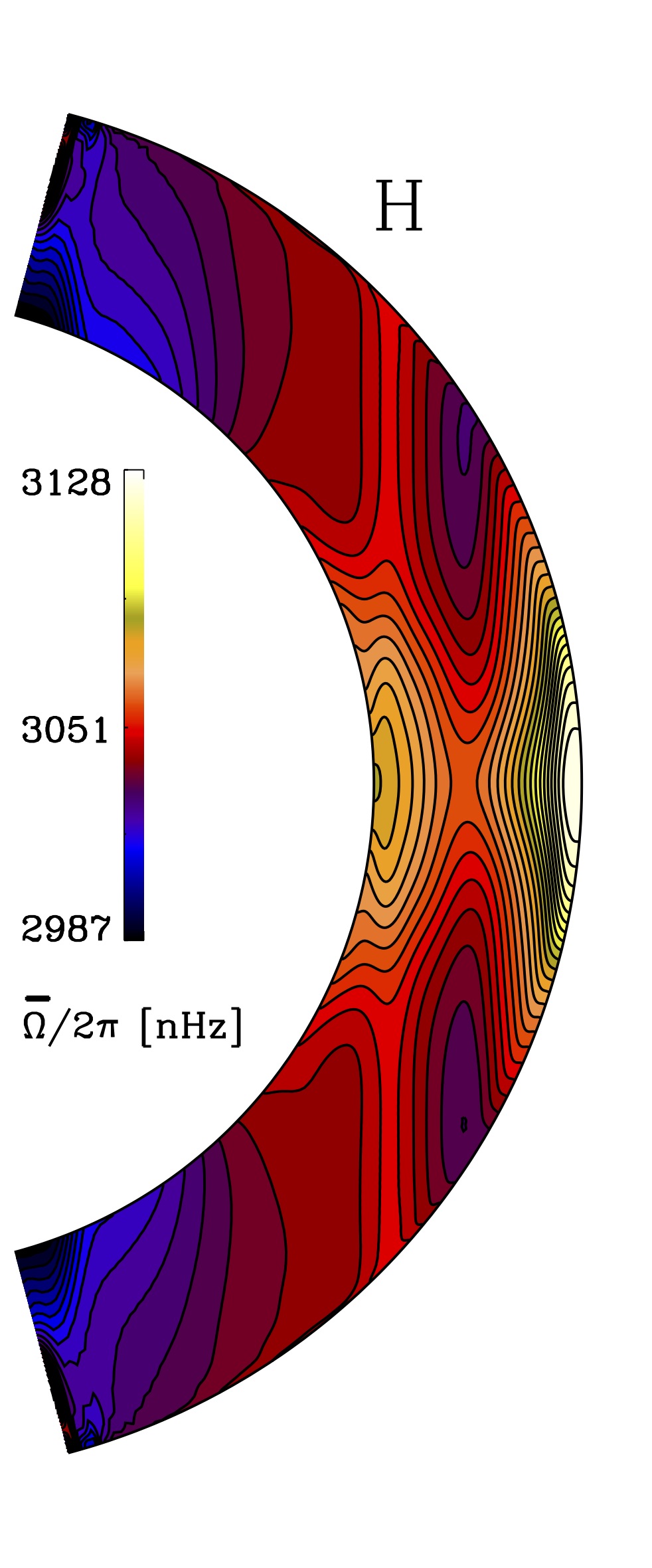}
\caption{Differential rotation profiles for runs C3, D, G$^a$, H from
\cite{VWKKOCLB18}, and MHD2 from \cite{KVKBF19}.
The green lines for MHD2 indicate 
the internal structure as for the
simulations in the main text.}\label{fig:KrComp}
\end{figure*}

We reproduce in Figure~\ref{fig:KrComp}, for comparison, the
differential rotation profiles for the simulations with fixed heat
conduction profiles (Runs~C3, D, G$^a$ and H) from \cite{VWKKOCLB18},
corresponding, respectively, to R1, R2 and R4 (the latter for G$^a$ 
and H) presented in this paper. 
We also show again the wedge simulation MHD2 from \cite{KVKBF19},
this time showing both the hemispheres.

The rotational profile of Run~D 
was not presented in the previous publications.
An equatorial prograde flow and mid-latitude minima are present, while in 
its Kramers counterpart, Run~R2, the profile is swapped to anti-solar with 
a decelerated equator and accelerated mid-latitude regions.

\bibliographystyle{aa}
\bibliography{bib}

\end{document}